\documentclass[useAMS,usenatbib,usegraphicx]{mn2e}
\usepackage[total={17.8cm,24.0cm},centering]{geometry}
\def\spose#1{\hbox to 0pt{#1\hss}}
\def\lesssim{\mathrel{\spose{\lower 3pt\hbox{$\sim$}}\raise 2.0pt\hbox{$<$}}}
\def\gtrsim{\mathrel{\spose{\lower 3pt\hbox{$\sim$}}\raise 2.0pt\hbox{$>$}}}
\def\kn{\hbox{$K\!$n}}
\title[$K$-band observations of boxy bulges.~I.]
 {$K$-band observations of boxy bulges.~I. Morphology and surface brightness profiles}
\author[M.\ Bureau et al.]
 {M.\ Bureau$^{1}$\thanks{E-mail: bureau@astro.ox.ac.uk}, G.\
 Aronica$^{2,3}$, E.\ Athanassoula$^3$, R.-J.\ Dettmar$^2$,
\newauthor A.\ Bosma$^3$, K.\ C.\ Freeman$^4$\\
$^1$Sub-Department of Astrophysics, University of Oxford, Denys
 Wilkinson Building, Keble Road, Oxford OX1~3RH, United Kingdom\\
$^2$Astronomisches Institut, Ruhr-Universit\"{a}t Bochum, D-44780 Bochum,
 Germany\\
$^3$Observatoire de Marseille, 2 place Le Verrier, F-13248 Marseille Cedex~4,
 France\\
$^4$Research School of Astronomy and Astrophysics, The Australian National
 University, Weston Creek P.O., ACT~2611, Australia}
\begin{document}
\maketitle
%
%
\begin{abstract}
In this first paper of a series on the structure of boxy and
peanut-shaped (B/PS) bulges, \kn-band observations of a sample of $30$
edge-on spiral galaxies are described and discussed. \kn-band
observations best trace the dominant luminous galactic mass and are
minimally affected by dust. Images, unsharp-masked images, as well as
major-axis and vertically-summed surface brightness profiles are
presented and discussed. Galaxies with a B/PS bulge tend to have a
more complex morphology than galaxies with other bulge types, more
often showing centered or off-centered X structures, secondary maxima
along the major-axis and spiral-like structures. While probably not
uniquely related to bars, those features are observed in
three-dimensional N-body simulations of barred discs and may trace the
main bar orbit families. The surface brightness profiles of galaxies
with a B/PS bulge are also more complex, typically containing $3$ or
more clearly separated regions, including a shallow or flat
intermediate region (Freeman Type~II profiles). The breaks in the
profiles offer evidence for bar-driven transfer of angular momentum
and radial redistribution of material. The profiles further suggest a
rapid variation of the scaleheight of the disc material, contrary to
conventional wisdom but again as expected from the vertical resonances
and instabilities present in barred discs. Interestingly, the steep
inner region of the surface brightness profiles is often shorter than
the isophotally thick part of the galaxies, itself always shorter than
the flat intermediate region of the profiles. The steep inner region
is also much more prominent along the major-axis than in the
vertically-summed profiles. Similarly to other recent work but
contrary to the standard `bulge + disc' model (where the bulge is both
thick and steep), we thus propose that galaxies with a B/PS bulge are
composed of a thin concentrated disc (a disc-like bulge) contained
within a partially thick bar (the B/PS bulge), itself contained within
a thin outer disc. The inner disc likely formed secularly through
bar-driven processes and is responsible for the steep inner region of
the surface brightness profiles, traditionally associated with a
classic bulge, while the bar is responsible for the flat intermediate
region of the surface brightness profiles and the thick complex
morphological structures observed. Those components are strongly
coupled dynamically and are formed mostly of the same (disc) material,
shaped by the weak but relentless action of the bar resonances. Any
competing formation scenario for galaxies with a B/PS bulge, which
represent at least $45$~per cent of the local disc galaxy population,
must explain equally well and self-consistently the above
morphological and photometric properties, the complex gas and stellar
kinematics observed, and the correlations between them.
\end{abstract}
%
%
\section{INTRODUCTION}
\label{sec:intro}
Since the work of \citet{defis83}, bulges have largely been considered
as low-luminosity elliptical galaxies, suggesting a rapid formation
dominated either by mergers and accretion of external material (e.g.\
\citealt*{k96,kcw96}; but see also \citealt*{bc01,abp01}) or possibly
by dissipative gravitational collapse (e.g.\
\citealt*{els62,c84a,c84b,sm95}; but see also \citealt{n98,n99}). Over
the last decade, however, there has been much criticism of this
idea. In particular, the realization that most bulges have an inner
surface brightness profile shallower than the expected $R^{1/4}$ law
\citep*[e.g.][]{apb95,j96,mch03,bgdp03} argues against both mechanisms
\citep[e.g.][]{hm95,abp01}. Alternative models where bulges grow
secularly (i.e.\ over a long timescale and in relative isolation) have
also been developed and studied, many of them bar-driven
\citep*[e.g.][]{pn90,fb93,fb95,nsh96}, and much observational data
support them \citep[e.g.][]{wfmmb95,es02}.

Of primary concern here, several pieces of evidence point to the
identification of most boxy and peanut-shaped (B/PS) bulges in edge-on
spiral galaxies with the bars of barred spirals. In $N$-body
simulations, whenever a disc galaxy forms a bar, a B/PS bar/bulge
develops soon after. This process was studied first by \citet{cs81},
later on by \citet{cdfp90} and \citet{rsjk91}, and more recently by
\citet{mwhdb95}, \citet{am02}, \citet{a02,a03,a05} and \citet*{msh05}.

The observed incidence of B/PS bulges is consistent with that expected
if they are associated with relatively strong bars. Recent work by
\citet*{ldp00a,ldp00b} demonstrates that $45$ per cent of all bulges
are B/PS, while amongst those the exact shape of the bulge depends
mainly on the viewing angle to the bar. As shown by the numerical
simulations, true peanuts are bars seen side-on, i.e.\ with the
major-axis of the bar roughly perpendicular to the line-of-sight. For
less favourable viewing angles, the bulge/bar looks boxy, and if the
bar is seen end-on it looks almost spherical. Stronger bars also lead
to more prominent peanut shapes, as demonstrated observationally
\citep[e.g.][]{ldp00b} and theoretically \cite[e.g.][]{am02,ba05}.

The kinematics of discs harboring a B/PS bulge, as measured from both
ionized-gas emission lines and stellar absorption lines, show the
behaviour expected of barred spirals viewed edge-on. This has been
demonstrated by the observations of \citet{km95}, \citet{mk99},
\citet{bf97,bf99} and \citet{cb04}, and by the modeling of
\citet{ab99} and \citet{ba99,ba05}. Similar tests are now also
available for face-on bars \citep{dcmm05}.

There is thus ample evidence that edge-on galaxies with a B/PS bulge
are simply barred disc galaxies, and that the B/PS bulges themselves
represent the thickest parts of the bars (see \citealt{a05} for a
review of all arguments). Yet there has been little convincing
evidence for this from surface photometry alone, the best work being
that of \citet{ldp00a,ldp00b}. Early studies of non-spheroidal bulges
used mostly optical images \citep[e.g.][]{j87,sg89,s93} and the
interpretation was often hampered by the large amount of
extinction. As we will show in this paper, it truly takes the
combination of N-body simulations and orbit studies with near-infrared
(NIR) images to derive direct and convincing photometric evidence
relating B/PS bulges and bars.

\citet*{spa02a,spa02b} and \citet*{psa02,psa03a} studied the orbital
structure of three-dimensional (3D) bars exhaustively \citep[but see
also][]{p84,pf91}. They find families of orbits which can not only
provide the backbone of the boxy and peanut shapes, but can also cause
local enhancements within the disc itself. Since extinction is far
less important in the NIR, $K$-band images are the ideal tool to study
the morphology of galaxies with a B/PS bulge, to look for similarities
with barred orbital structures.

In this paper, the first of a series, we study a sample of $30$
edge-on spiral galaxies, most of them with a B/PS bulge, for which we
have obtained high-quality $K$-band images. Much complementary data
exist for this sample \citep[e.g.][]{bf97,bf99,cb04,bc06}, but the
primary goal here is to study the morphology of the B/PS structures
and their host discs, similarly to \citet{ldp00b}. As we shall see, we
find several features ressembling closely those expected of 3D bars:
bulges with X shapes, secondary disc enhancements, inner rings, etc. A
second paper discusses scaleheight variations in the same galaxies
(\citealt*{aab06}, hereafter \citeauthor{aab06}; but see also
\citealt{aabbdvp03,abad04}).

We present our sample in \S~\ref{sec:sample}, discuss the observations
and data reduction in \S~\ref{sec:obs}, and then describe and discuss
the resulting images and surface brightness profiles in
\S~\ref{sec:images} and \ref{sec:sbprofs}, respectively. We examine
the direct consequences of our results in \S~\ref{sec:discussion} and
conclude briefly in \S~\ref{sec:conclusions}.
%
%
\section{SAMPLE DESCRIPTION}
\label{sec:sample}
We study here the sample of \citet{bf99} and \citet{cb04}, which was
drawn from the catalogs of galaxies with a B/PS bulge of \citet{j86},
\citet{s87} and \citet{sa87}, as well as from galaxies with extreme
axial ratios ($a/b\ge7$; \citealt*{kkp93}). We include galaxies with
extended as well as no or confined ionized-gas emission, even though
the latter were not specifically studied by \citet{bf99}. The sample
thus consists of $30$ edge-on spirals, all of which are visible from
the south ($\delta\lesssim15^\circ$) and have a bulge larger than
$0\farcm6$ in diameter (to be able to identify bars kinematically in
moderate seeing). $24$ of the bulges were classified as having a B/PS
bulge by \citet{bf99} and $6$ constitute a `control' sample and have
varied morphologies.

Basic galaxy properties are listed in Table~\ref{tab:sample}, along
with a classification of the bulges' shape from
\citet{ldp00a}. Although still relying on a visual inspection of
optical contour plots, \citet{ldp00a} homogeneously classified over
$1350$ edge-on galaxies, including all our objects. Only $4$ bulges
were classified differently by \citet{bf99} and \citet{ldp00a}. A
complete comparison of the classifications as well as of the
ionized-gas and stellar kinematics is provided in
\citet{cb04}. \citet{bc06} discuss in more depth $3$ objects where
counter-rotating ionized-gas was discovered \citep[see
also][]{ckbg06}. As demonstrated by \citet{ldp00a}, the classification
of the bulges' shape is robust between the optical and NIR, but NIR
images are essential to study their detailed structure, due to the
frequent and significant extinction from dust in the equatorial plane
\citep[e.g.][]{ldp00b}. All sample galaxies therefore also have
$D_{25}\lesssim7\arcmin$ (where $D_{25}$ is the diameter at the
$25$~mag~arcsec$^{-2}$ isophotal level in $B$), which allows for the
NIR imaging discussed here to be acquired in a reasonable amount of
time with a small-field camera.
%
%
\begin{table*}
\begin{minipage}{177mm}
\caption{Galaxy sample.}
\label{tab:sample}
\begin{tabular}{lrrlrrrrr}
\hline
Galaxy & R.A.(2000) & Dec.(2000) & Type & Bulge & $B_{\rm T}$ &
$M_B^{\rm c}$ & $D_{25}$ & $V_{\odot}$~~~ \\
 & $^{~h~~~m~~~s}$ & $^{~~~\degr~~~\arcmin~~~\arcsec}$ & & & ~mag &
~~~~mag & arcmin & km~s$^{-1}$ \\
\multicolumn{1}{c}{(1)} & \multicolumn{1}{c}{(2)} & \multicolumn{1}{c}{(3)} &
  \multicolumn{1}{c}{(4)} & \multicolumn{1}{c}{(5)} & \multicolumn{1}{c}{(6)}
  & \multicolumn{1}{c}{(7)} & \multicolumn{1}{c}{(8)} & \multicolumn{1}{c}{(9)}\\
\hline
\noalign{\vspace{0.15cm}}
\multicolumn{9}{l}{\bf B/PS bulges} \\
NGC 128      & 00 29 15.1 & +02 51 50 & S0 pec       & 1.0 & 12.7 & -21.4 & 2.81 & 4228 \\ 
ESO 151-G004 & 00 56 07.3 & -53 11 28 & S0$^{0}$     & 1.0 & 14.7 & -20.4 & 1.31 & 7456 \\
NGC 1381     & 03 36 31.7 & -35 17 43 & SA0          & 2.0 & 12.7 & -19.2 & 2.63 & 1776 \\
NGC 1596     & 04 27 37.8 & -55 01 37 & SA0          & 4.0 & 12.0 & -19.3 & 3.89 & 1509 \\
NGC 1886     & 05 21 48.2 & -23 48 36 & Sab          & 1.0 & 13.8 & -19.2 & 3.23 & 1737 \\
NGC 2310     & 06 53 53.8 & -40 51 46 & S0           & 2.0 & 12.6 & -18.6 & 4.16 & 1187 \\
ESO 311-G012 & 07 47 34.0 & -41 27 07 & S0/a?        & 2.0 & 12.4 & -20.0 & 3.71 & 1130 \\
NGC 2788A    & 09 02 40.2 & -68 13 38 & Sb           & 1.0 & 13.6 & -21.5 & 2.88 & 4056 \\
IC  2531     & 09 59 55.4 & -29 37 02 & Sb           & 1.0 & 12.9 & -21.6 & 6.76 & 2474 \\
NGC 3203     & 10 19 34.4 & -26 41 53 & SA(r)0$^{+}$?& 3.0 & 13.0 & -19.9 & 2.81 & 2410 \\
NGC 3390     & 10 48 04.4 & -31 32 02 & Sb           & 2.0 & 12.8 & -21.5 & 3.46 & 3039 \\
NGC 4469     & 12 29 28.0 & +08 44 59 & SB(s)0/a?    & 1.0 & 12.4 & -17.8 & 3.46 &  576 \\
NGC 4710     & 12 49 38.9 & +15 09 57 & SA(r)0$^{+}$ & 1.5 & 11.9 & -19.8 & 4.89 & 1324 \\
PGC 44931    & 13 01 49.5 & -08 20 10 & Sbc          & 1.0 & 14.2 & -21.1 & 2.81 & 3804 \\
ESO 443-G042 & 13 03 29.9 & -29 49 36 & Sb           & 1.0 & 13.9 & -20.6 & 2.88 & 2912 \\ 
NGC 5746     & 14 44 55.9 & +01 57 17 & SAB(rs)b?    & 1.0 & 11.4 & -21.8 & 6.91 & 1720 \\
IC  4767     & 18 47 41.6 & -63 24 20 & S pec        & 1.0 & 14.3 & -19.5 & 1.51 & 3544 \\
NGC 6722     & 19 03 39.6 & -64 53 41 & Sb           & 1.0 & 13.5 & -22.2 & 2.88 & 5749 \\
NGC 6771     & 19 18 39.6 & -60 32 46 & SA(r)0$^{+}$?& 1.0 & 13.6 & -20.5 & 2.34 & 4221 \\
ESO 185-G053 & 20 03 00.4 & -55 56 53 & SB pec       & 2.0 & 14.3 & -20.0 & 1.23 & 4475 \\
IC  4937     & 20 05 17.9 & -56 15 20 & Sb           & 1.0 & 14.8 & -18.6 & 1.86 & 2337 \\
ESO 597-G036 & 20 48 15.0 & -19 50 58 & S0$^{0}$ pec & 1.0 & 15.2 & -20.7 & 0.87 & 8694 \\
IC  5096     & 21 18 21.8 & -63 45 42 & Sb           & 4.0 & 13.6 & -20.7 & 3.16 & 3142 \\
ESO 240-G011 & 23 37 50.5 & -47 43 37 & Sb           & 4.0 & 13.4 & -21.0 & 4.89 & 2842 \\
\hline
\noalign{\vspace{0.15cm}}
\multicolumn{9}{l}{\bf Control sample}\\
NGC 1032     & 02 39 23.6 & +01 05 38 & S0/a         & 4.0 & 12.7 & -20.7 & 3.46 & 2722 \\
NGC 3957     & 11 54 01.5 & -19 34 09 & SA0$^{+}$    & 3.0 & 13.0 & -19.0 & 3.09 & 1686 \\
NGC 4703     & 12 49 18.9 & -09 06 30 & Sb           & 4.0 & 14.0 & -21.1 & 2.45 & 4458 \\
NGC 5084     & 13 20 16.8 & -21 49 38 & S0           & 4.0 & 11.5 & -20.9 &10.71 & 1725 \\
NGC 7123     & 21 50 46.4 & -70 19 59 & Sa           & 4.0 & 13.6 & -20.3 & 2.51 & 3737 \\
IC  5176     & 22 14 55.3 & -66 50 56 & SAB(s)bc?    & 4.0 & 13.4 & -19.6 & 4.36 & 1746 \\
\hline
\end{tabular}
\\
Notes: All parameters are from LEDA (Lyon-Meudon Extragalactic
Database), except the morphological type \citep{j86,sa87,s87,kkp93}
and bulge type \citep{ldp00a}. Bulge types $1$--$3$ represent B/PS
bulges and bulge type $4$ spheroidal bulges. ESO~151-G004's redshift
and corrected absolute $B$ magnitude are from \citet{cb04}.
\end{minipage}
\end{table*}
%
%
\section{OBSERVATIONS AND DATA REDUCTION}
\label{sec:obs}
The observations were carried out at the 2.3m telescope of Siding
Spring Observatory over $8$ runs from 1996 January to 1997 March, for
a total of $24$ nights. Mounted at the $f/18$ Cassegrain focus, the
CASPIR\footnote{http://www.mso.anu.edu.au/observing/2.3m/CASPIR/}
(Cryogenic Array Spectrometer/Imager; \citealt{mhdhb94}) instrument
was used in direct imaging mode with a $0.5$~arcsec~pix$^{-1}$
scale. Its $256\times256$~pix$^2$ SBRC InSb CCD ($30\micron$ pixels)
then yields a $128$~arcsec~$\times$~$128$~arcsec field-of-view. All
observations were obtained with a narrow $K$-band filter, \kn, of
central wavelength $2.165\micron$ and width $0.33\micron$. The readout
noise with the double sample readout method used (relative sampling)
was about $60$~e and the dark current $\lesssim30$~e~s$^{-1}$ over
most of the array, although there are a number of hot pixels. For
$5$~s integrations (see below) and typical background brightness
conditions ($12.0$--$13.0$~\kn\ mag~arcsec$^{-2}$), our observations
are always background limited.

For all observations except biases and standard stars, the integration
time for a single frame was $60$~s, resulting from the average of
twelve $5$~s exposures (individual exposures were not saved). The
total integrations were built-up using a script taking $5$ frames of
each field, each dithered by $4\arcsec$ ($8$ pixels; the biggest bad
pixel patch) parallel or perpendicular to the galaxy major-axis, and
interspersed by sky frames. The lot was bracketed by dark and bias
frames and repeated typically $7$ times, for a total on-source
integration time of $35$~min per field, slightly less on sky. For a
galaxy requiring mosaicing of $2$ fields, for example, the sequence
was: bias, dark, sky, field\_1 [dith 0], field\_2 [dith 0], sky,
field\_1 [dith 1], field\_2 [dith 1], sky, field\_1 [dith 2], \ldots
This ensures sufficiently fast sampling of the sky and allows to
average out bad pixels when the frames are combined. Using the
instrument rotator, the CCD rows were always aligned (and centered) on
the galaxy major-axis, facilitating mosaicing. Mosaics of $1$ to $5$
fields (with $25$ per cent overlap) were required depending on the
galaxy, and all mosaics include plenty of sky area around the target.

The data reduction was mostly carried out within the CASPIR IRAF
(Image Reduction and Analysis Facility) package, following as closely
as possible the procedures recommended in the CASPIR manual. Dark,
sky, and object frames were first bias-subtracted and linearized, and
the skies and objects dark-subtracted. Sky and object frames were then
flatfielded and the sky subtracted from the objects. All the frames of
a given object were then combined together with proper offsets to form
the final image. The bias and dark frames used for subtraction were
nightly (or fractions of a night if deemed necessary) medians of all
such frames. For sky subtraction, running medians of a few sky frames
were used, time-centered around the object frame being reduced. A flat
was obtained every night from the difference of domes frames taken
with an incandescent lamp on and off. For most objects, a mosaic was
created for each dither position (e.g.\ for field\_1 [dith 0] and
field\_2 [dith 0] in the example above) using blind offsets (the
offsets given to the telescope), and all such mosaics were then added
by registering compact sources (galaxy nucleus and foreground
stars). This procedure minimizes registering errors due to a drift of
the telescope pointing over time, and is required because a guide star
common to all fields of a mosaic could not be obtained for most
targets (the overlap of the $5\farcm2$ guider fields being minimal or
non-existent). A common background across the different fields of each
mosaic was imposed by subtracting the proper constant from each field
when necessary (imperfect sky subtraction).

A photometric zero-point on the Carter SAAO system \citep{cm95} was
derived nightly by observing IRIS (Infrared Imaging Spectrograph)
photometric standards throughout the night. The \kn-band atmospheric
extinction correction coefficient was also calculated each night by
following a few standard stars over a range of airmasses (typical
value $\approx0.10$~mag~airmass$^{-1}$). The galaxy images were also
corrected for Galactic extinction, although this is negligible for
most objects. When frames were obtained under less than photometric
conditions, their fluxes were first scaled to the level of photometric
ones before being combined. The final images generally have a
background level slightly different from zero, and this was again
corrected by subtracting a constant.
%
%
\section{IMAGES}
\label{sec:images}
\subsection{Morphological features}
\label{sec:morphology}
Figures~\ref{fig:bps} and \ref{fig:control} show, respectively, our
data for galaxies with a B/PS bulge and for those of the control
sample. Each panel contains, from top to bottom, first a Digitized Sky
Survey (DSS) image of the galaxy (for comparison), second our \kn-band
image and third an unsharp-masked \kn-band image (see below), all
spatially registered. The surface brightness profiles which follow
will be discussed in the next section (\S~\ref{sec:sbprofs}). As our
main interest here is the galaxy bulges, we only show the inner parts
of our mosaics, which generally extend up to and beyond the
large-scale discs.
%
%
\begin{figure*}
\begin{center}
\includegraphics[width=0.45\textwidth]{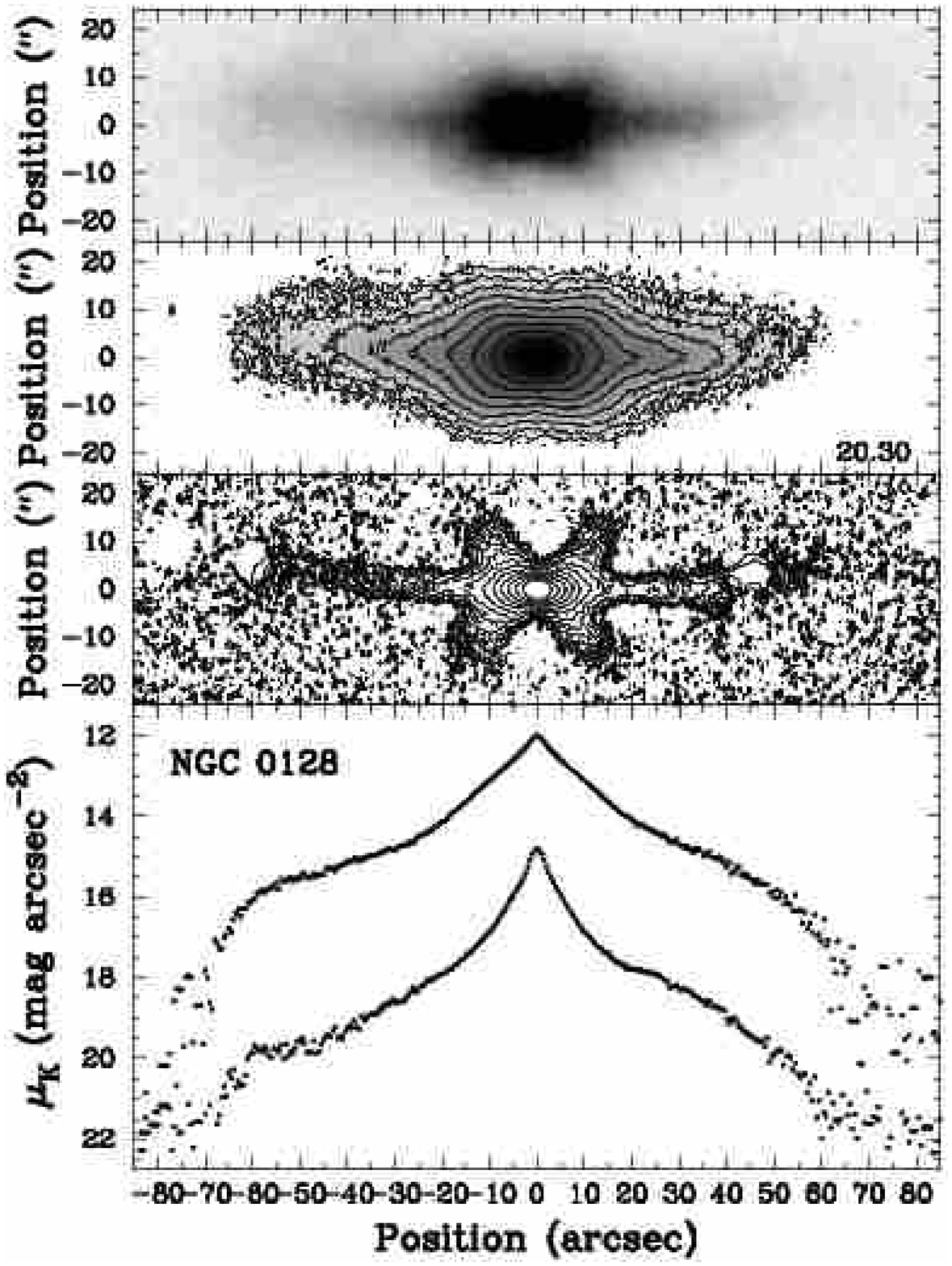}\hspace*{0.05\textwidth}\includegraphics[width=0.45\textwidth]{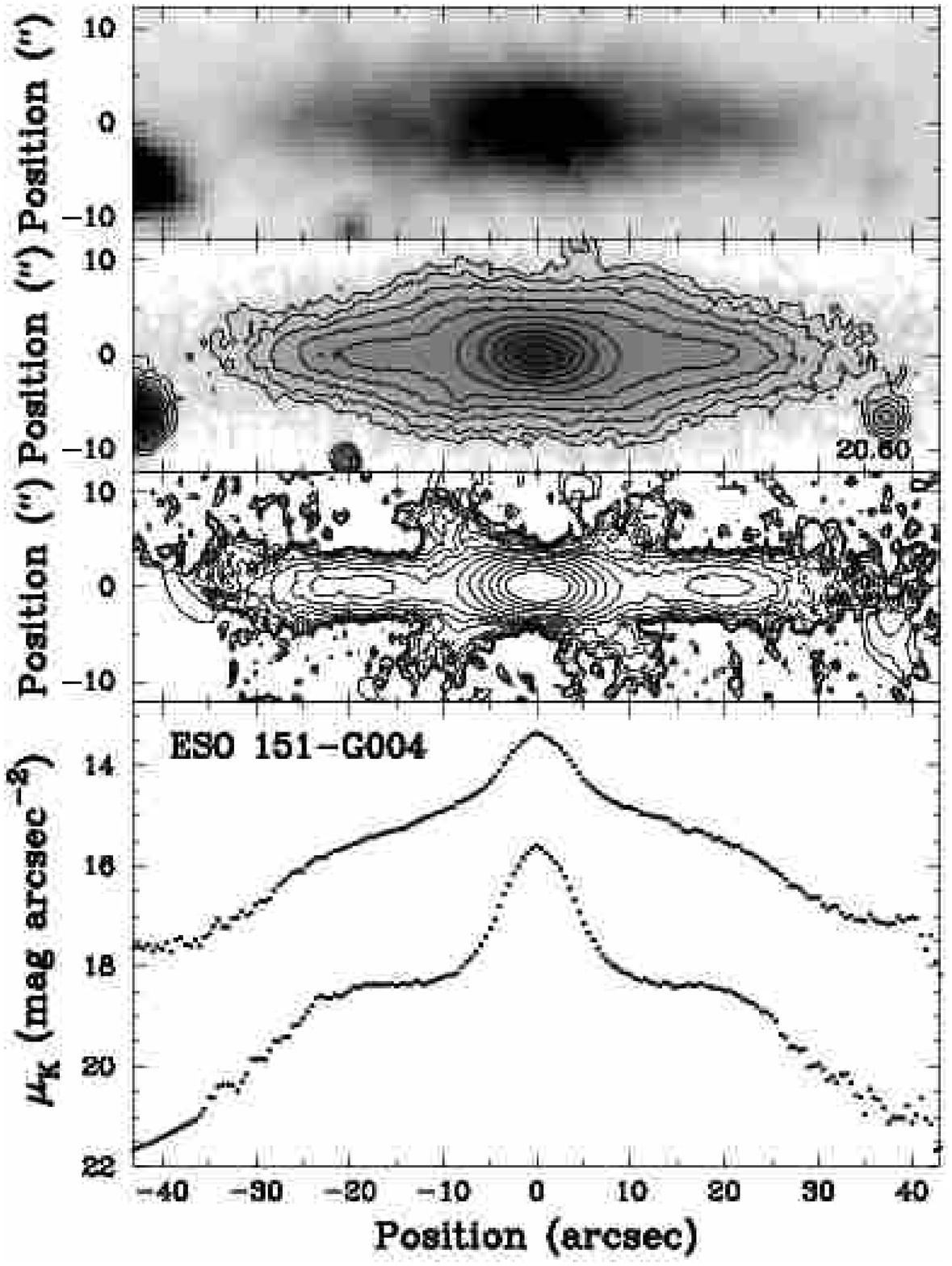}
\includegraphics[width=0.45\textwidth]{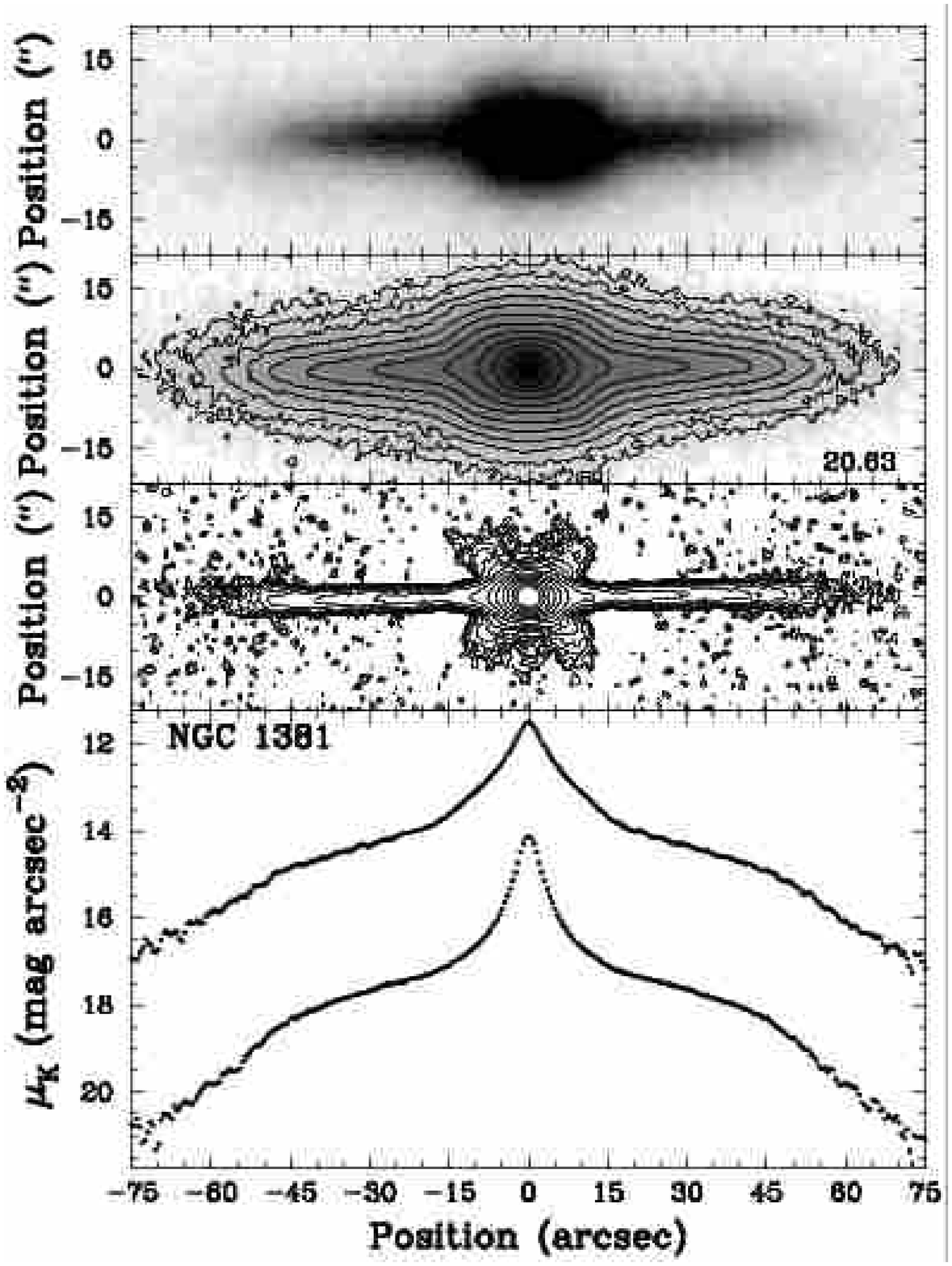}\hspace*{0.05\textwidth}\includegraphics[width=0.45\textwidth]{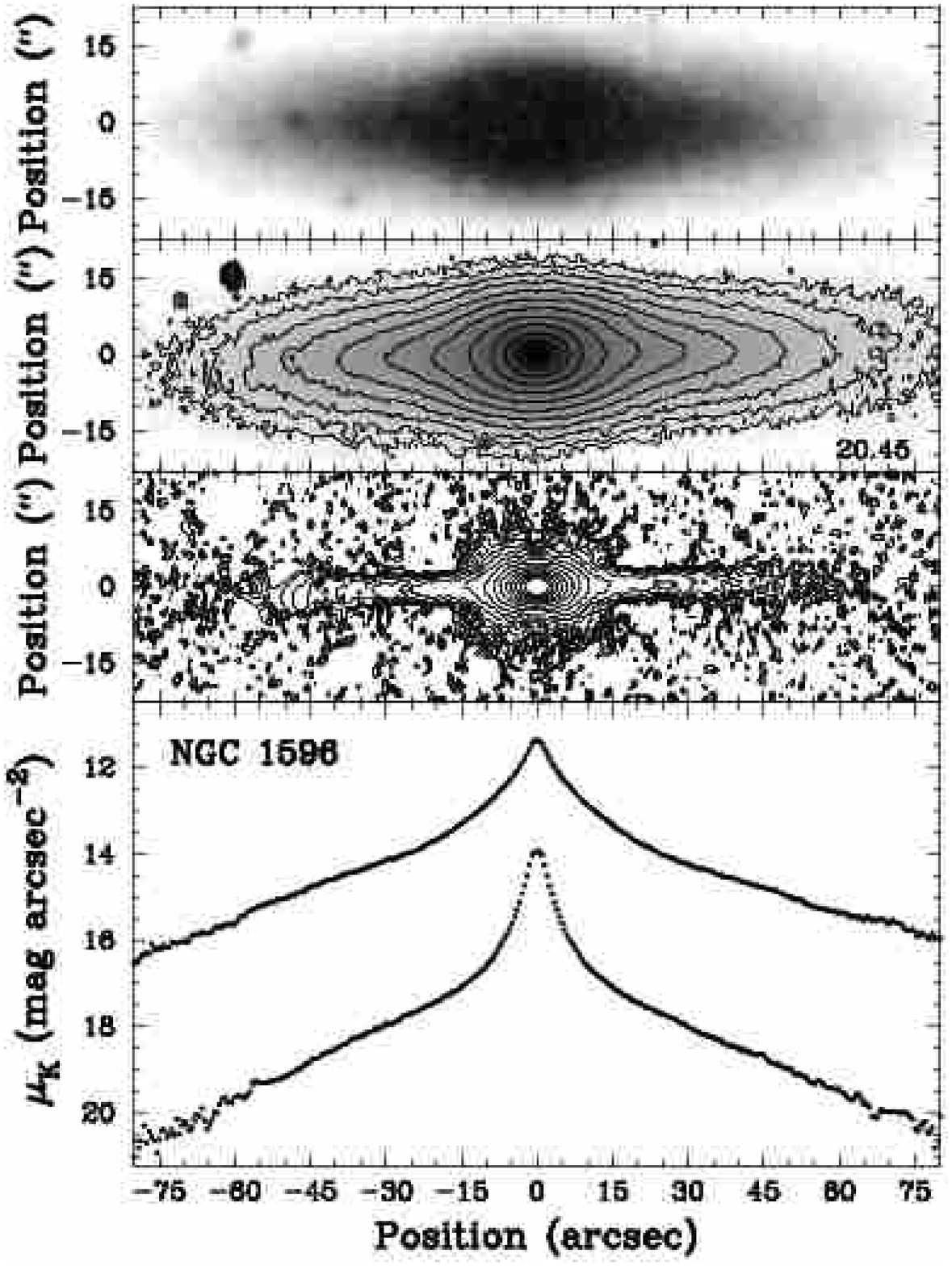}
\end{center}
\caption{Images and surface brightness profiles of the galaxies with a
 B/PS bulge. From top to bottom, each panel shows first a DSS image of
 the galaxy, second our \kn-band image, third an unsharp-masked
 \kn-band image and last major-axis (fainter) and summed (brighter)
 surface brightness profiles, all spatially registered. Contours for
 the \kn-band image are spaced by $0.5$ mag~arcsec$^{-2}$ and the
 faintest countour is indicated in the bottom-right corner of the
 panel.}
\label{fig:bps}
\end{figure*}
\begin{figure*}
\begin{center}
\includegraphics[width=0.45\textwidth]{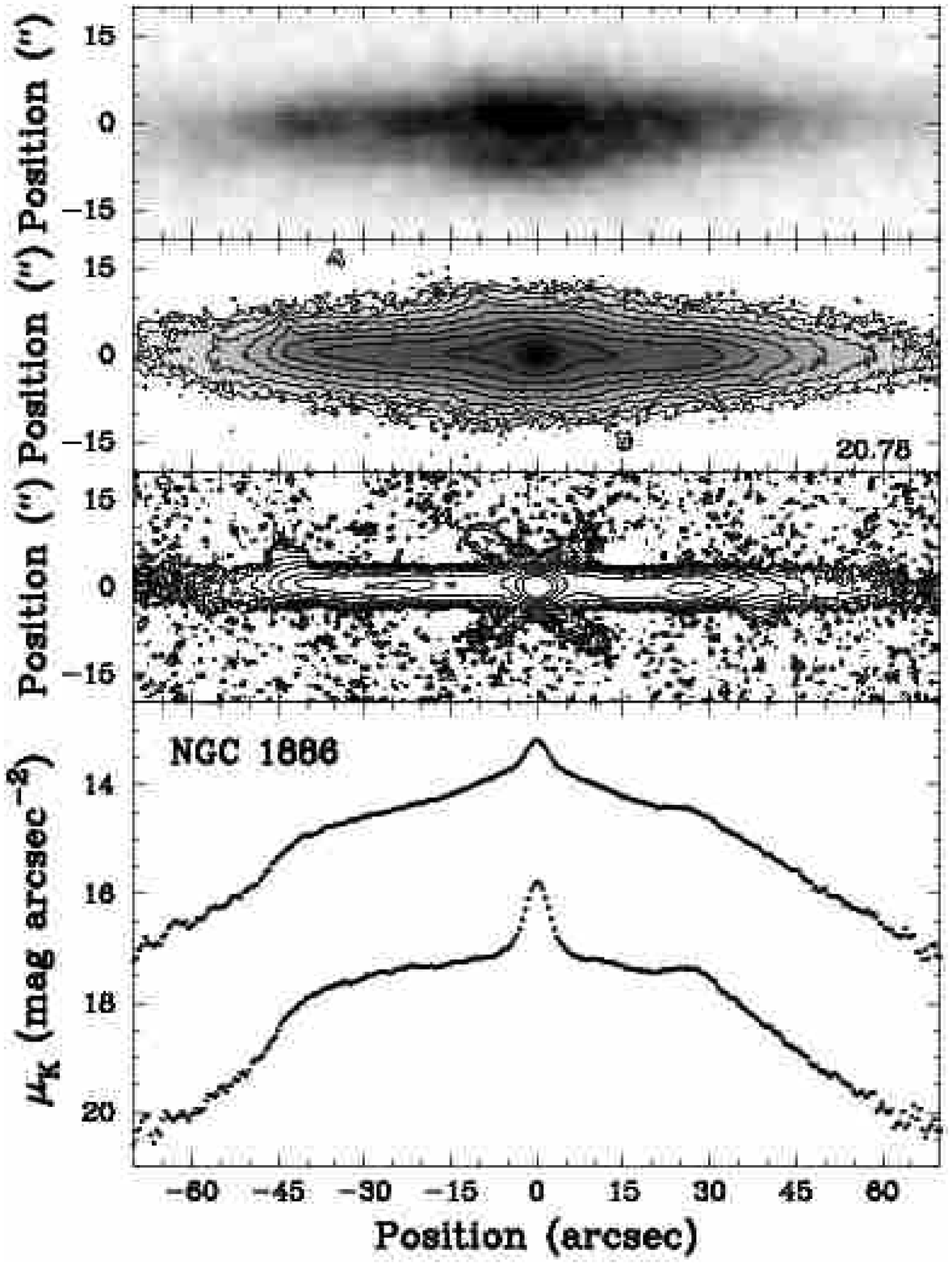}\hspace*{0.05\textwidth}\includegraphics[width=0.45\textwidth]{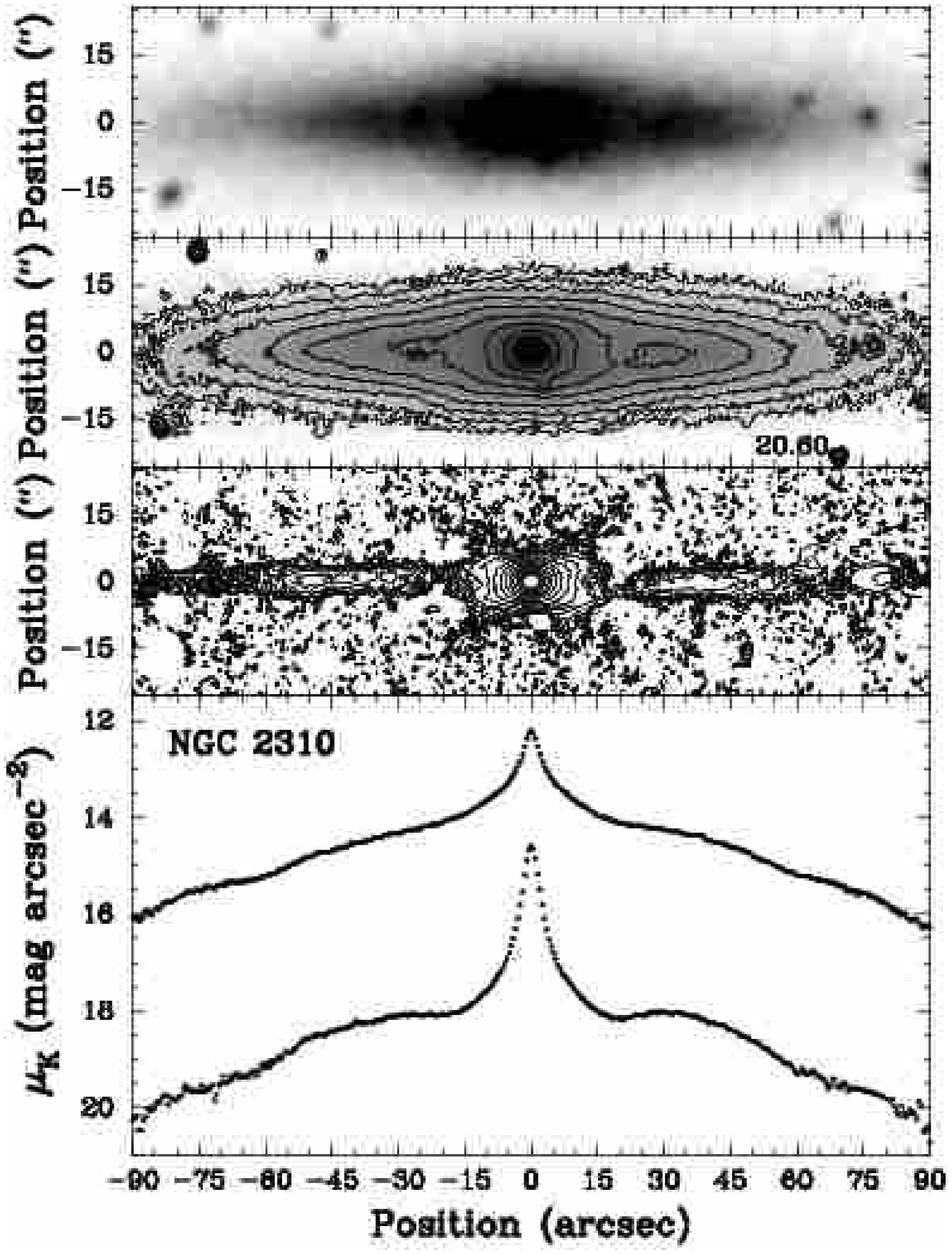}
\includegraphics[width=0.45\textwidth]{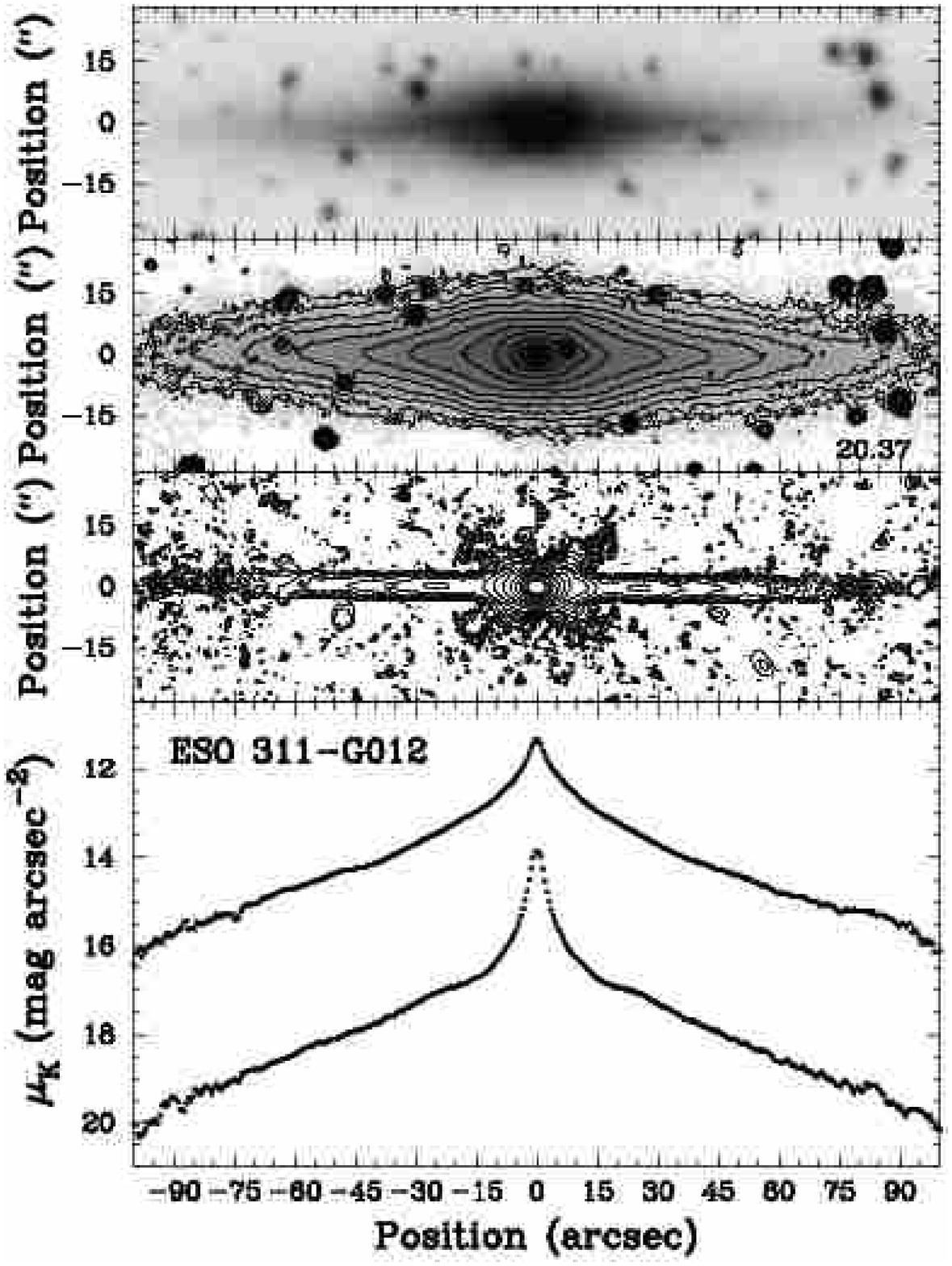}\hspace*{0.05\textwidth}\includegraphics[width=0.45\textwidth]{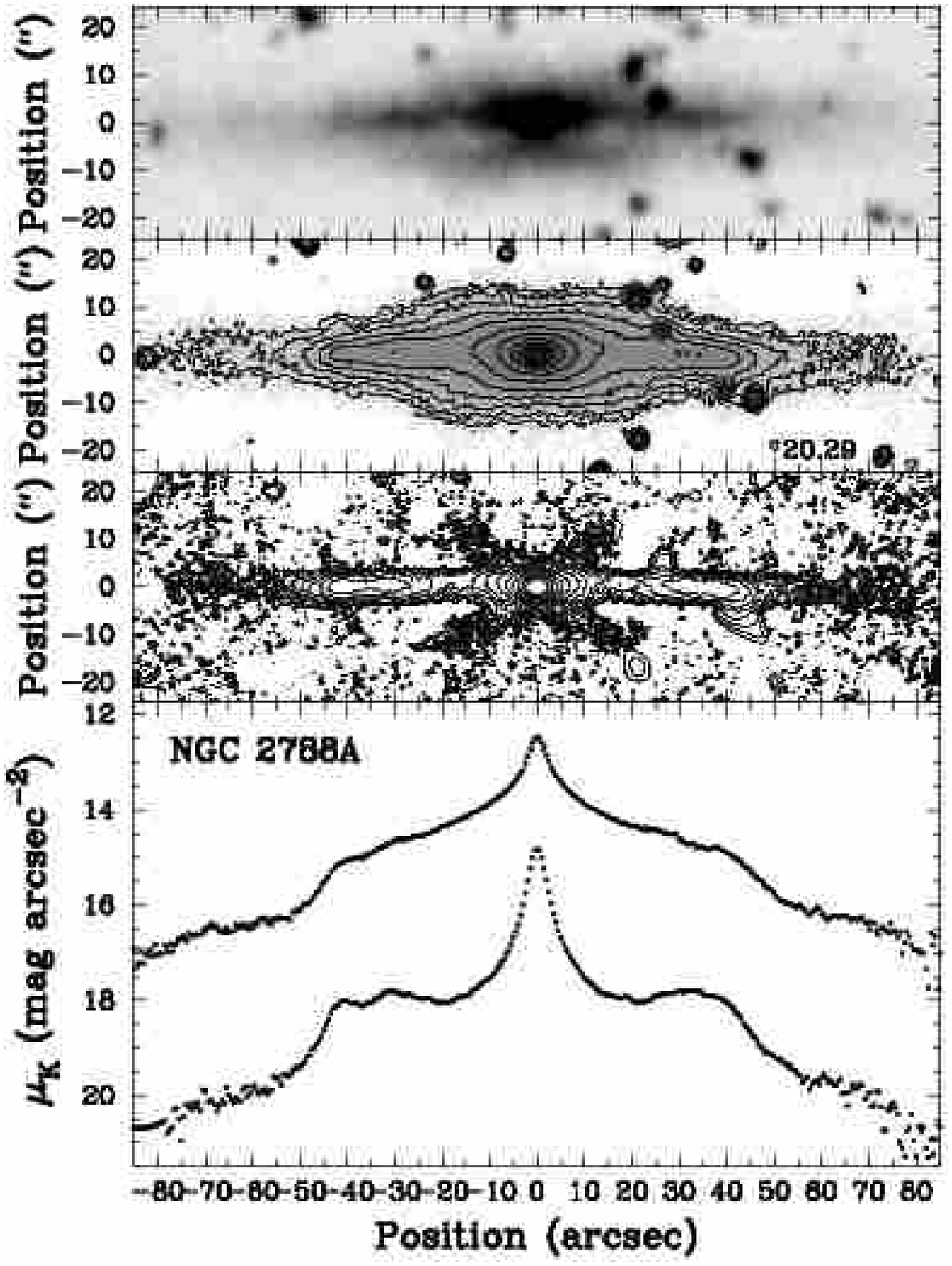}
\end{center}
\addtocounter{figure}{-1}
\caption{Continued.}
\end{figure*}
\begin{figure*}
\begin{center}
\includegraphics[width=0.45\textwidth]{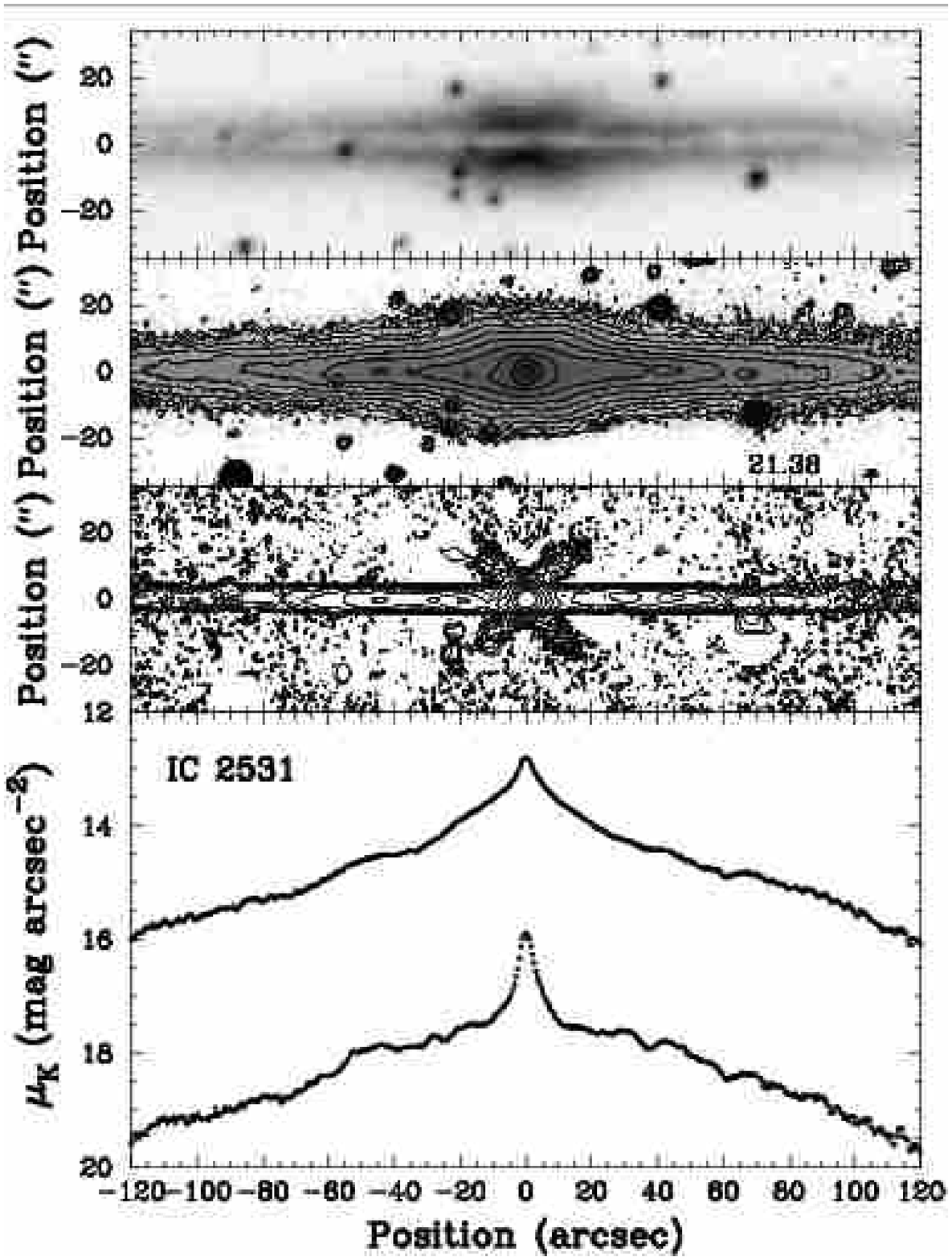}\hspace*{0.05\textwidth}\includegraphics[width=0.45\textwidth]{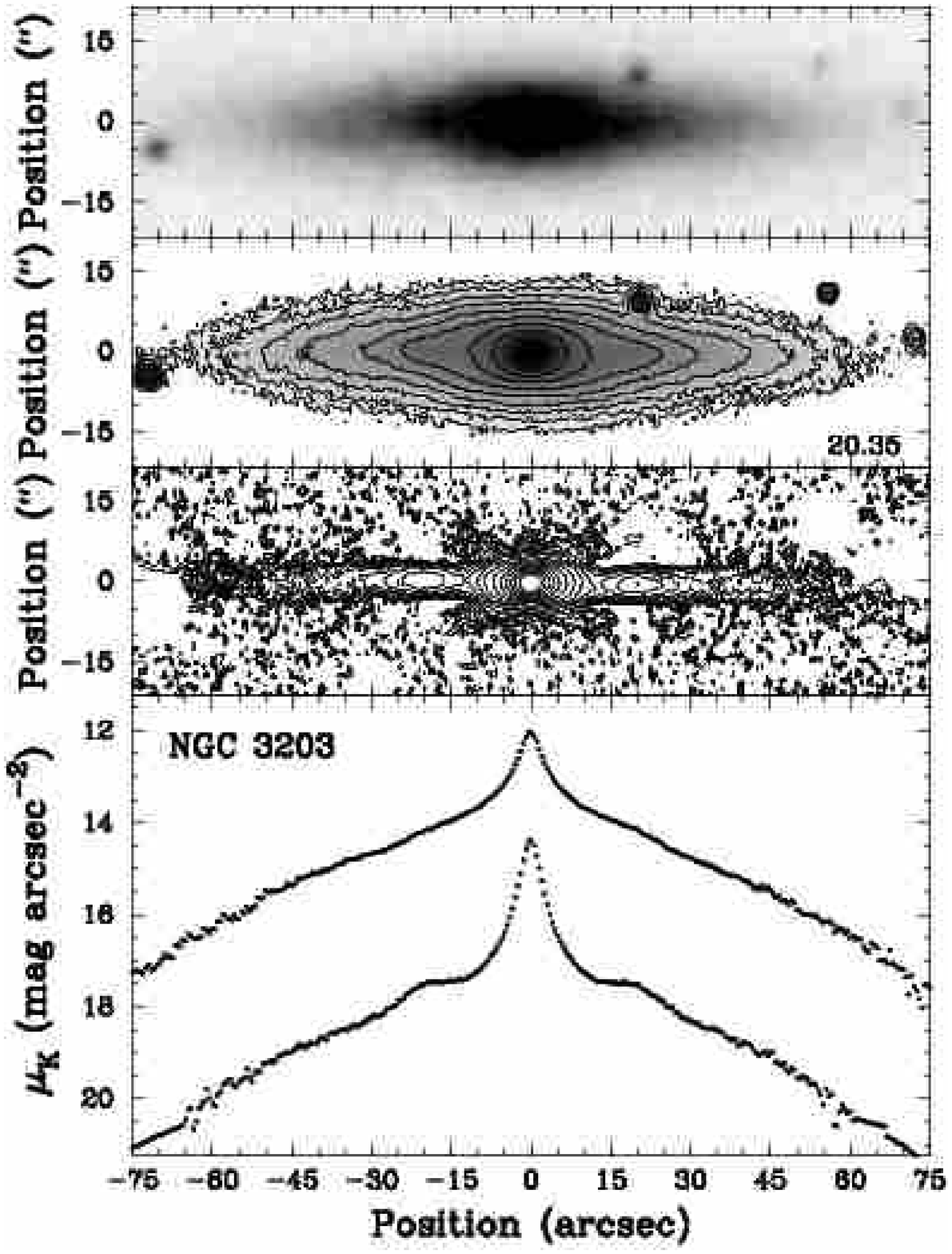}
\includegraphics[width=0.45\textwidth]{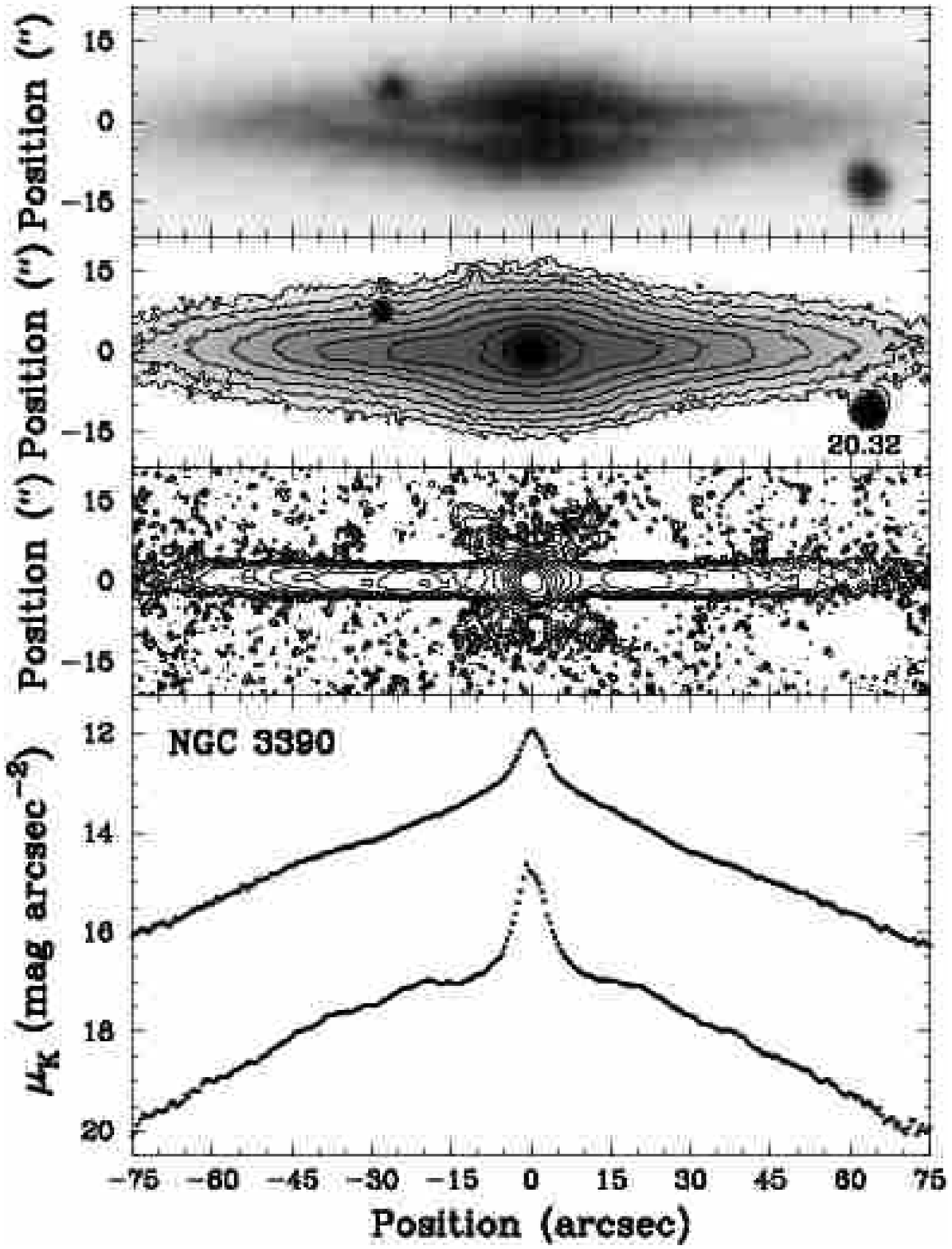}\hspace*{0.05\textwidth}\includegraphics[width=0.45\textwidth]{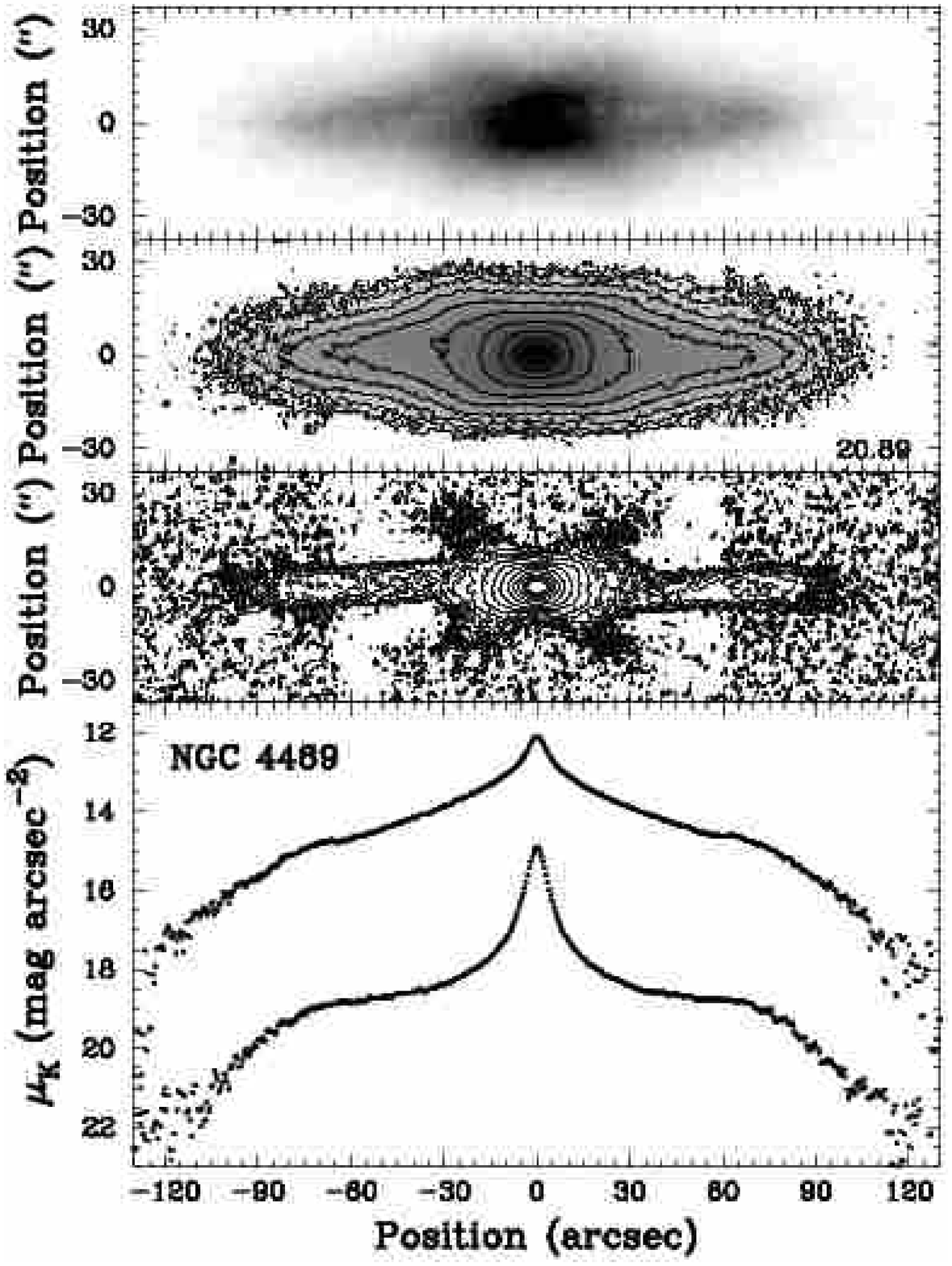}
\end{center}
\addtocounter{figure}{-1}
\caption{Continued.}
\end{figure*}
\begin{figure*}
\begin{center}
\includegraphics[width=0.45\textwidth]{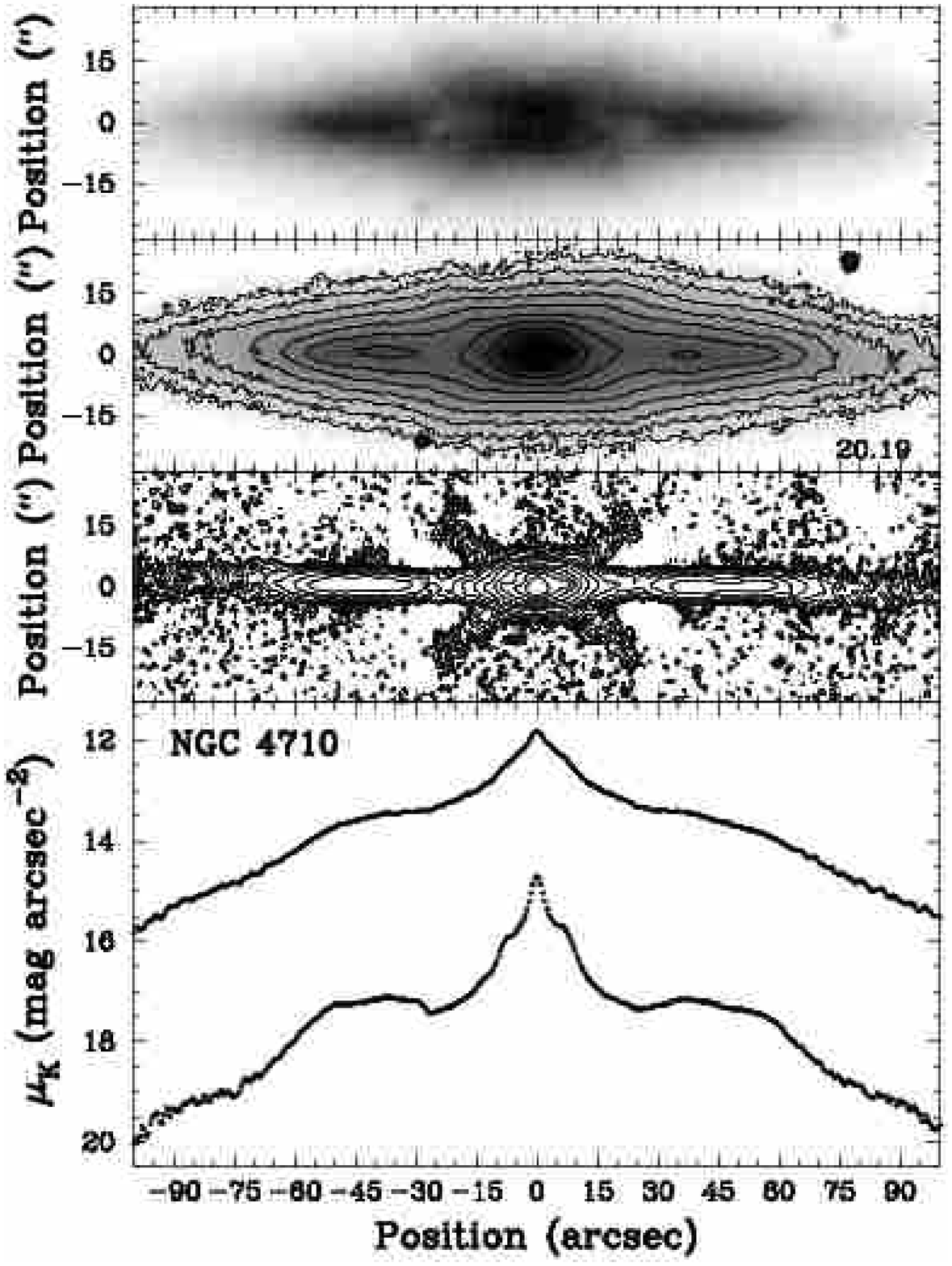}\hspace*{0.05\textwidth}\includegraphics[width=0.45\textwidth]{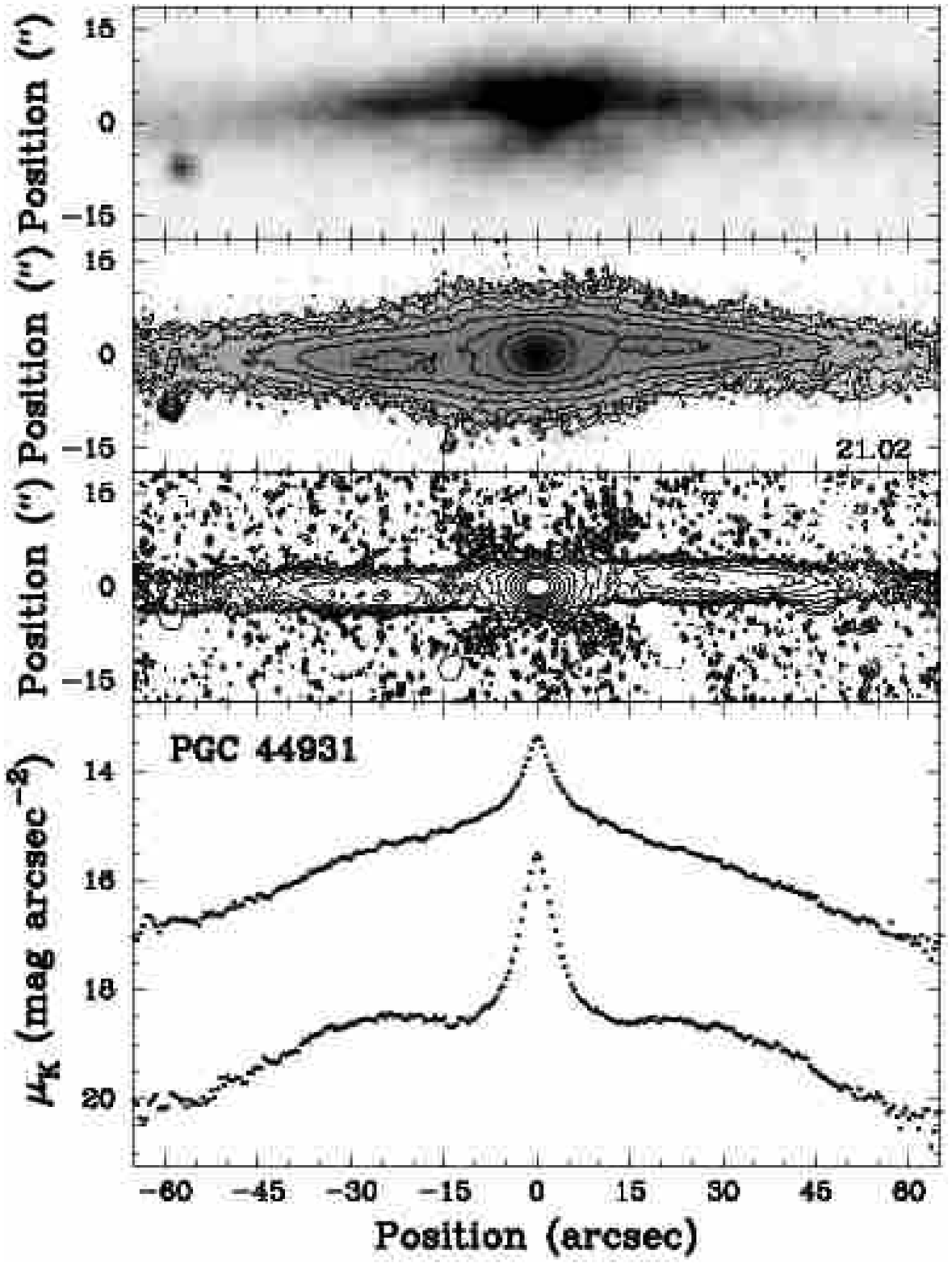}
\includegraphics[width=0.45\textwidth]{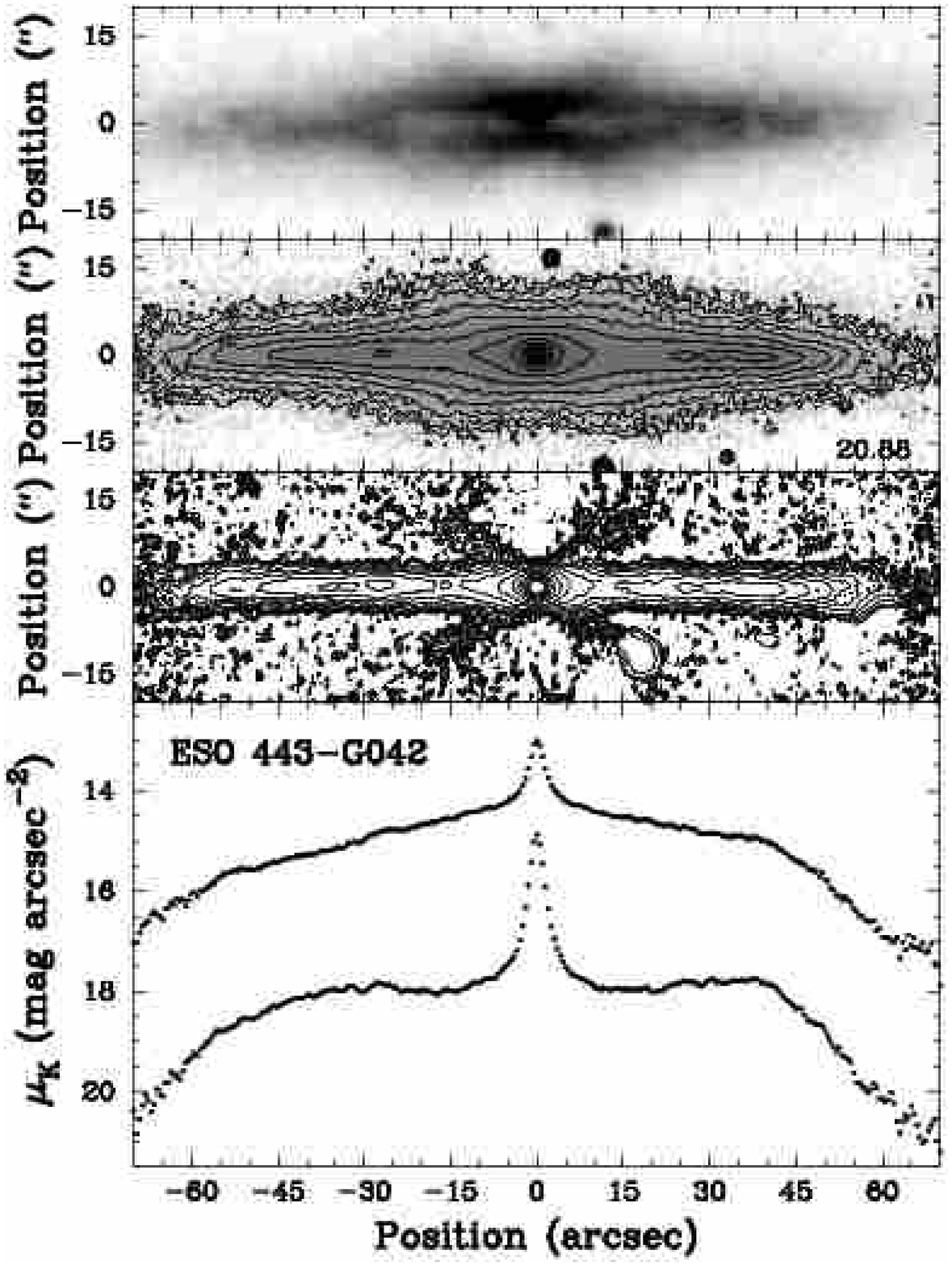}\hspace*{0.05\textwidth}\includegraphics[width=0.45\textwidth]{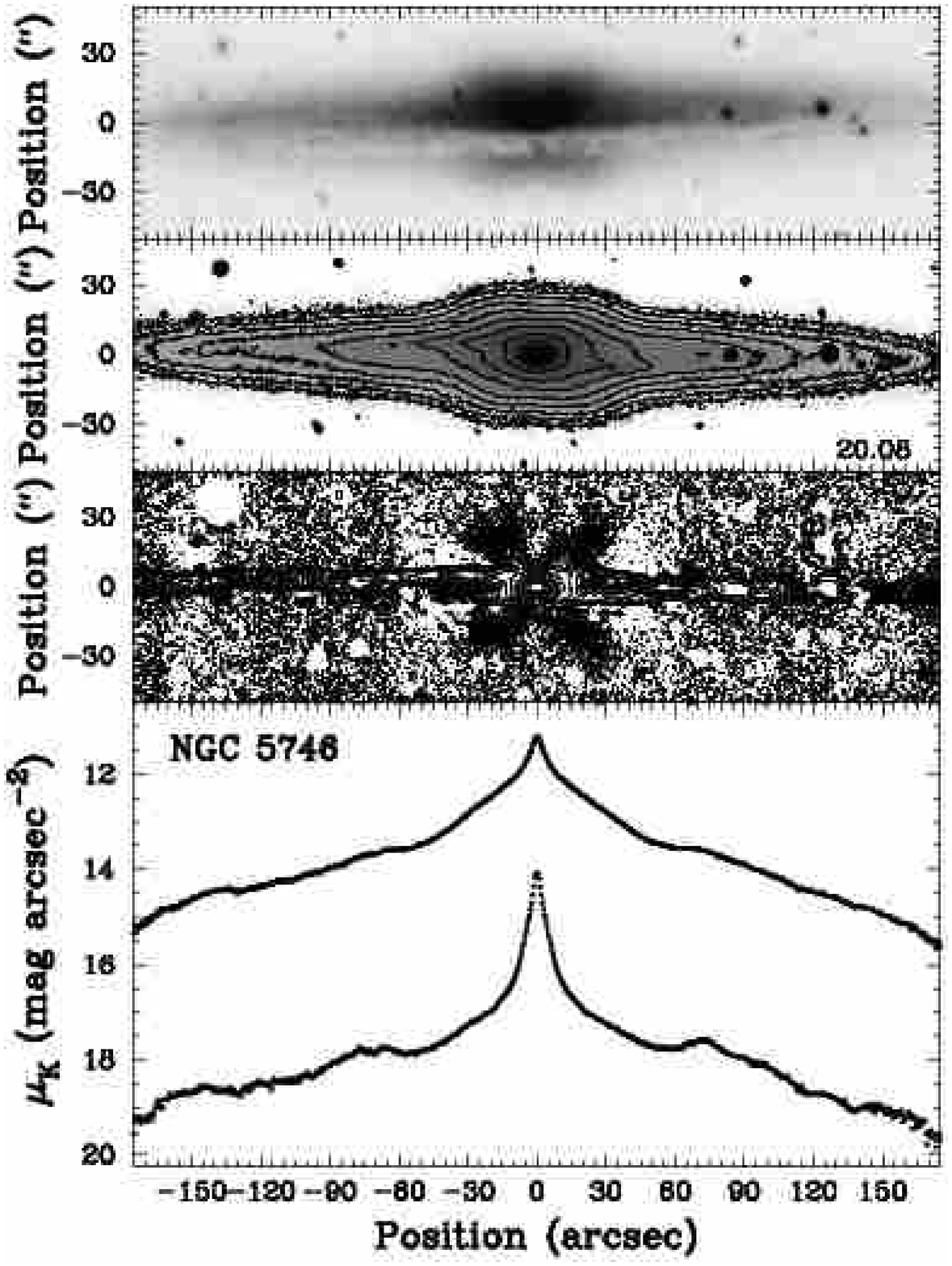}
\end{center}
\addtocounter{figure}{-1}
\caption{Continued.}
\end{figure*}
\begin{figure*}
\begin{center}
\includegraphics[width=0.45\textwidth]{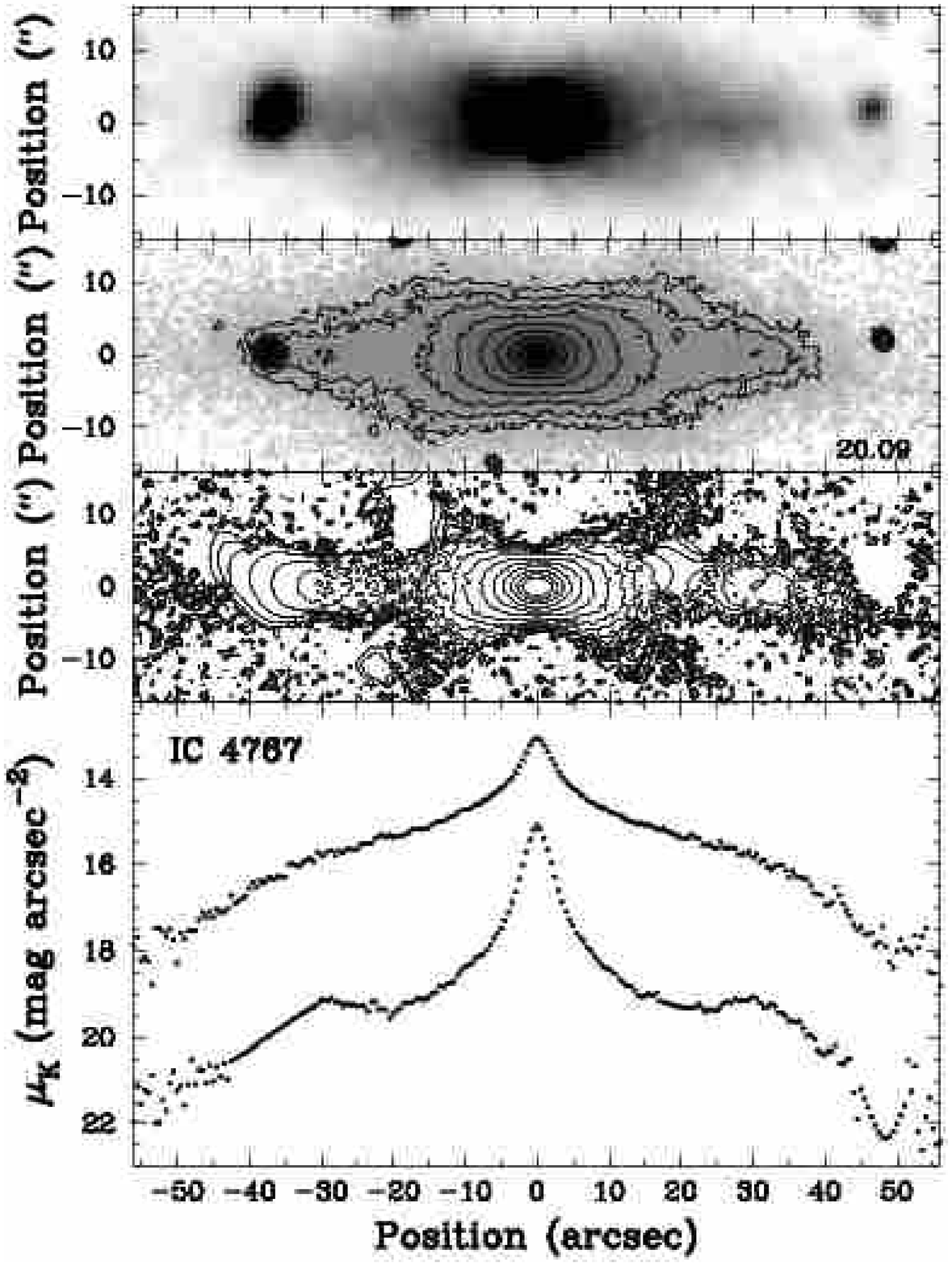}\hspace*{0.05\textwidth}\includegraphics[width=0.45\textwidth]{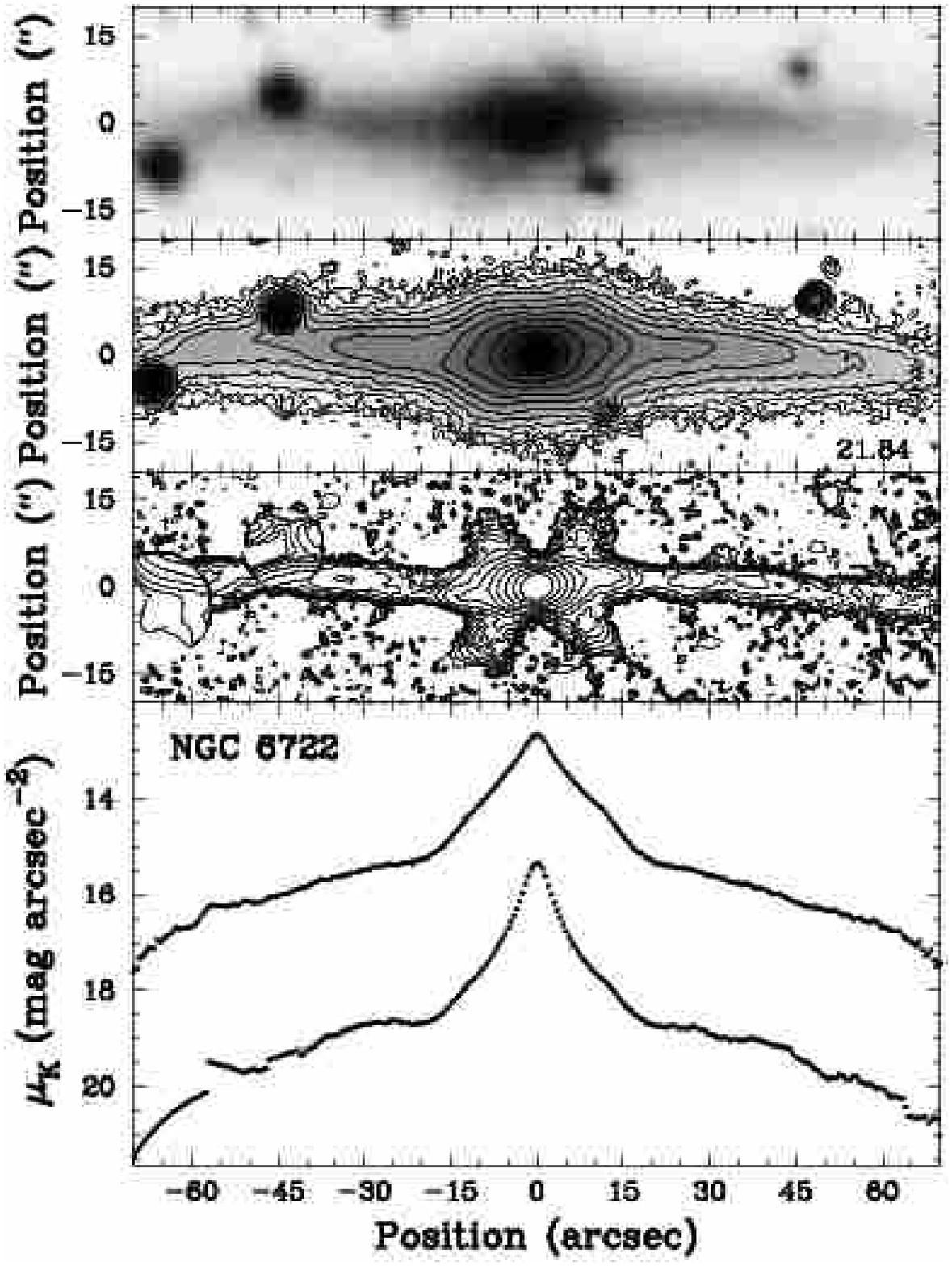}
\includegraphics[width=0.45\textwidth]{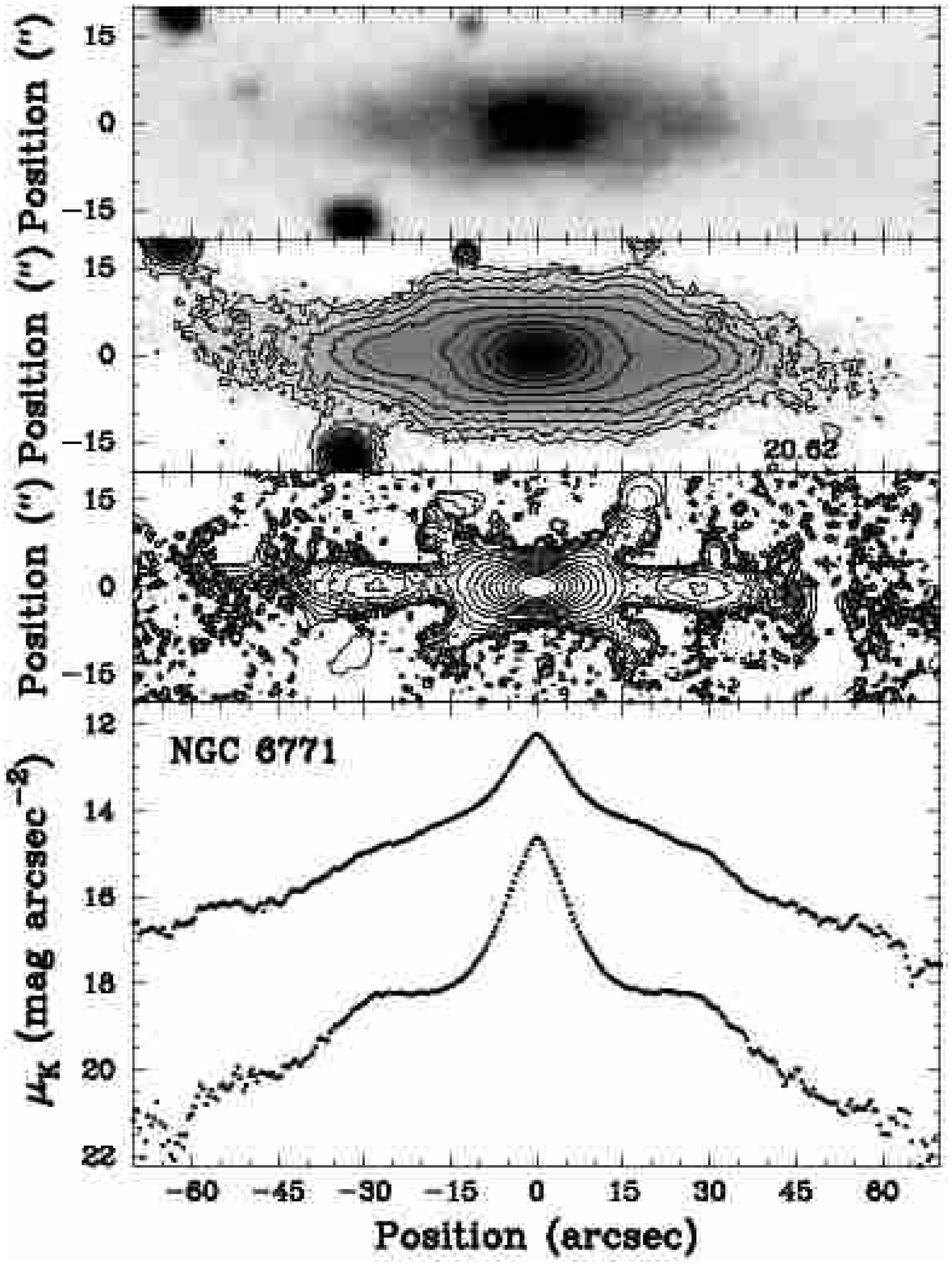}\hspace*{0.05\textwidth}\includegraphics[width=0.45\textwidth]{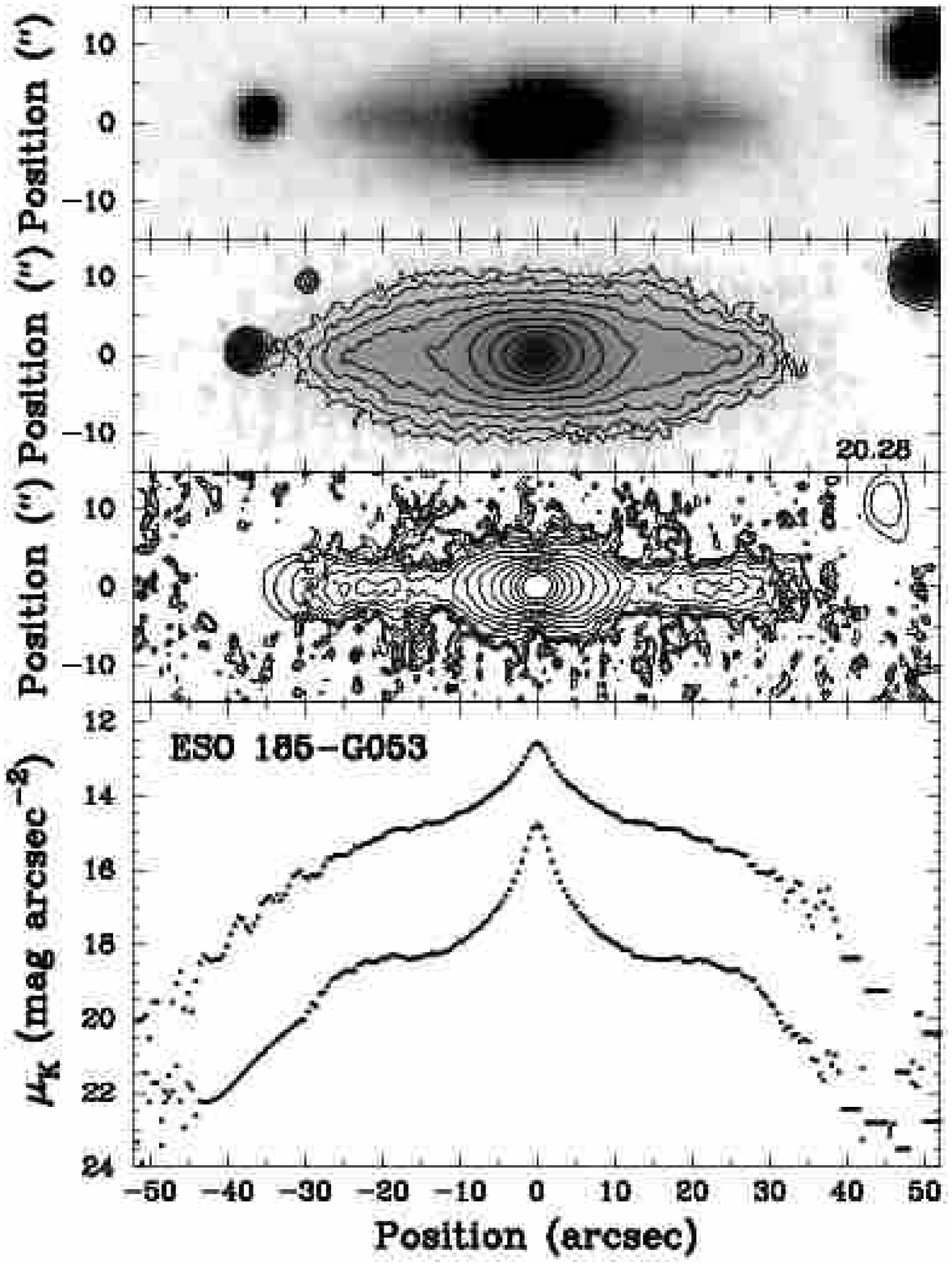}
\end{center}
\addtocounter{figure}{-1}
\caption{Continued.}
\end{figure*}
\begin{figure*}
\begin{center}
\includegraphics[width=0.45\textwidth]{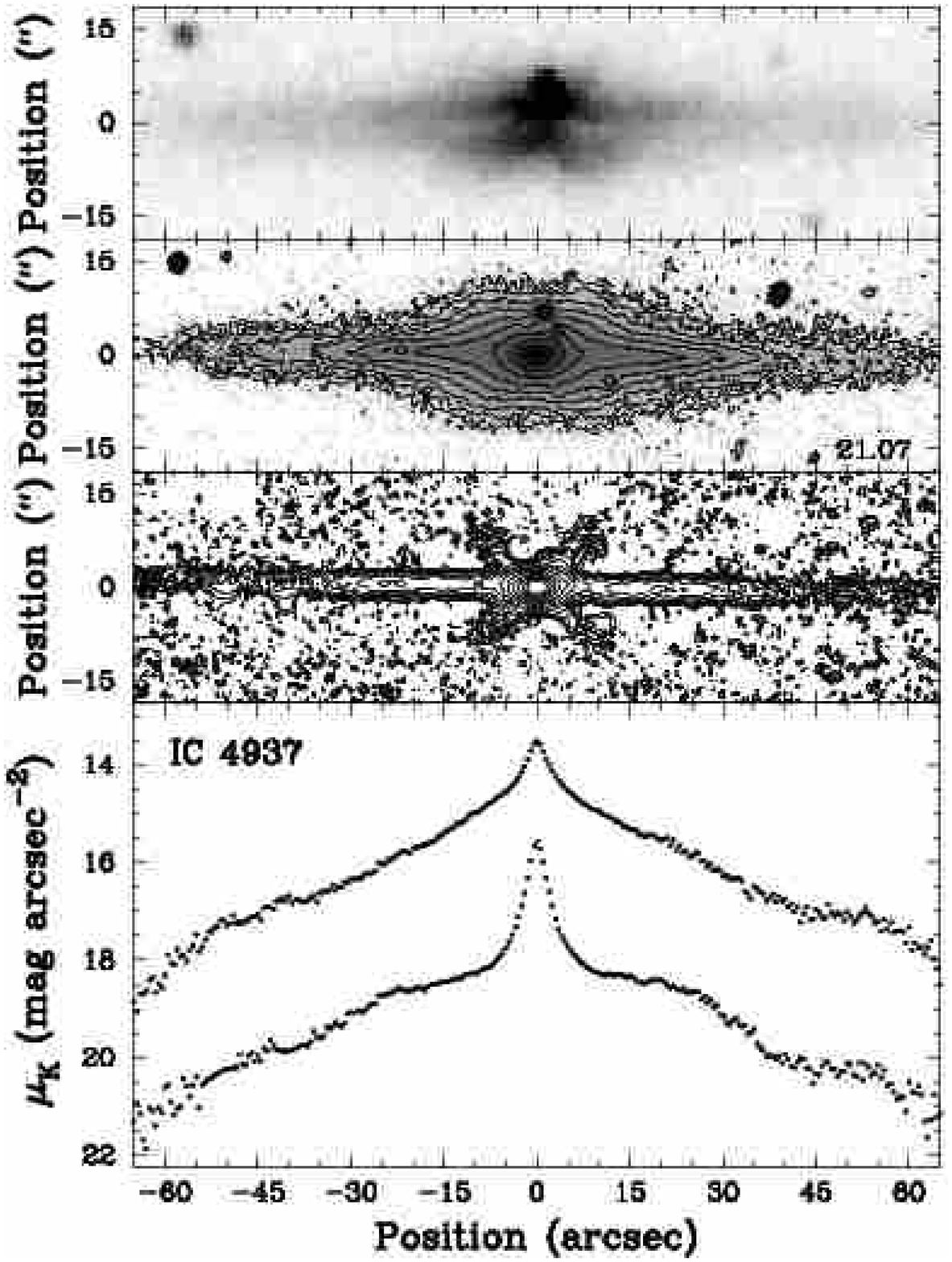}\hspace*{0.05\textwidth}\includegraphics[width=0.45\textwidth]{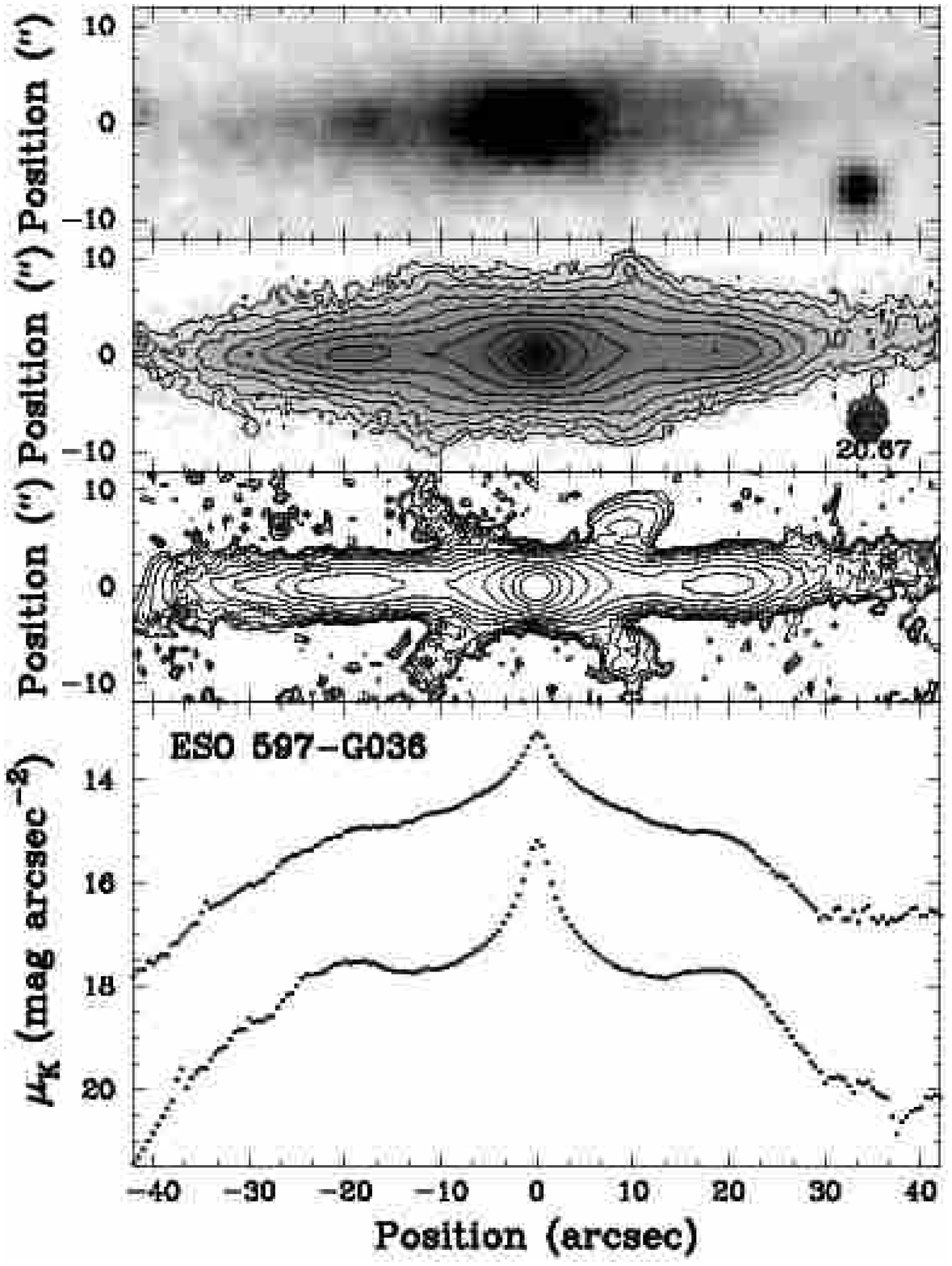}
\includegraphics[width=0.45\textwidth]{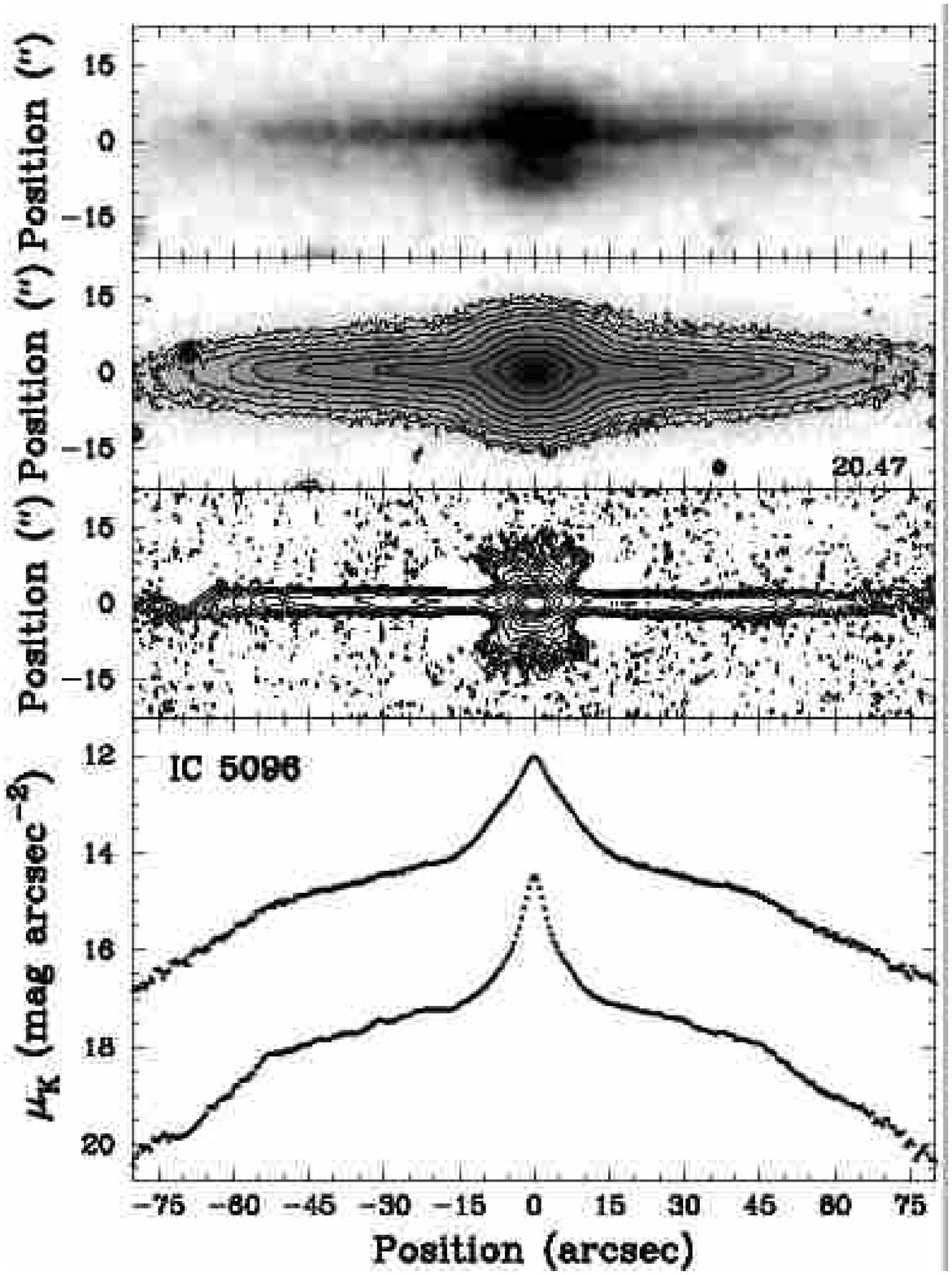}\hspace*{0.05\textwidth}\includegraphics[width=0.45\textwidth]{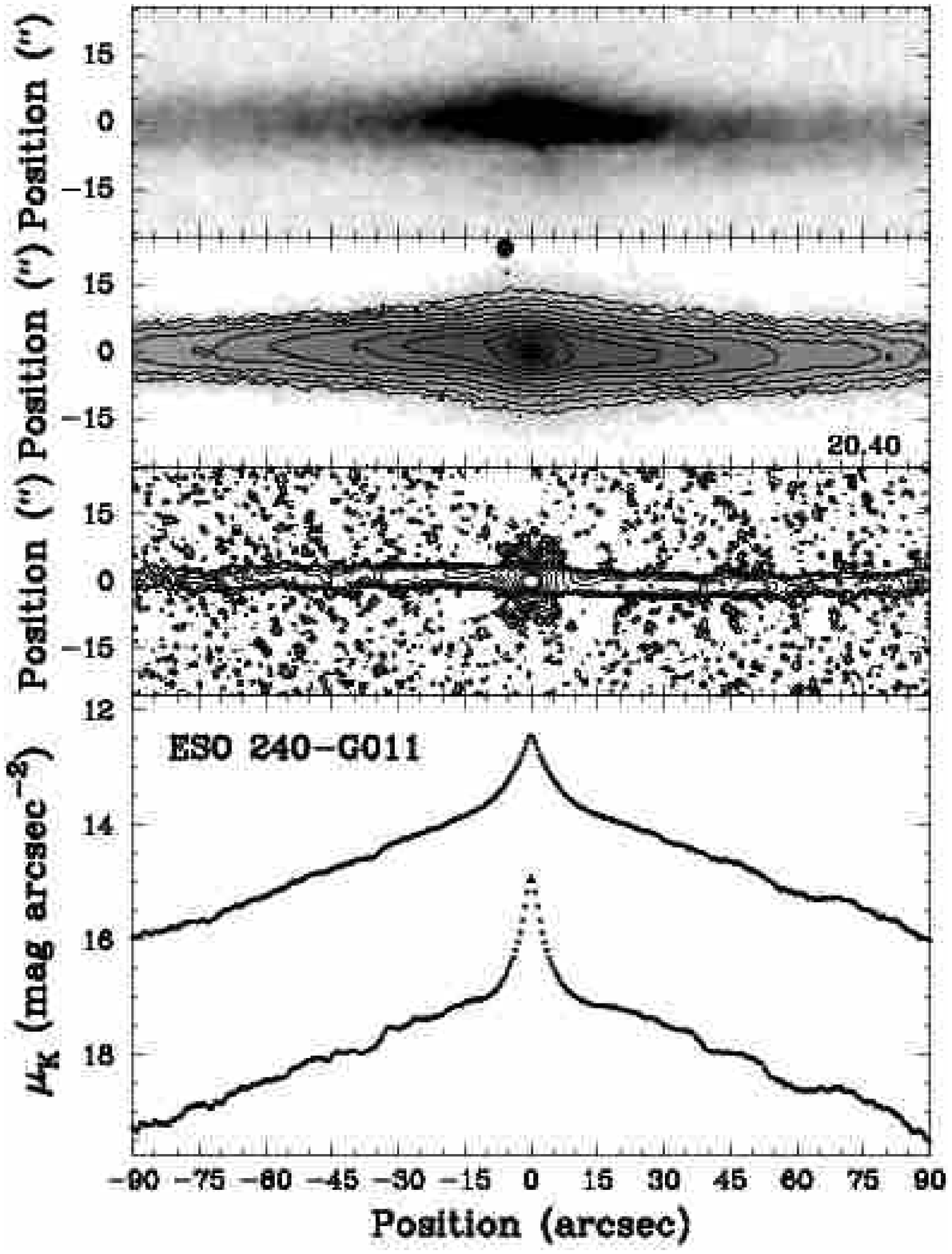}
\end{center}
\addtocounter{figure}{-1}
\caption{Continued.}
\end{figure*}
%
%
\begin{figure*}
\begin{center}
\includegraphics[width=0.45\textwidth]{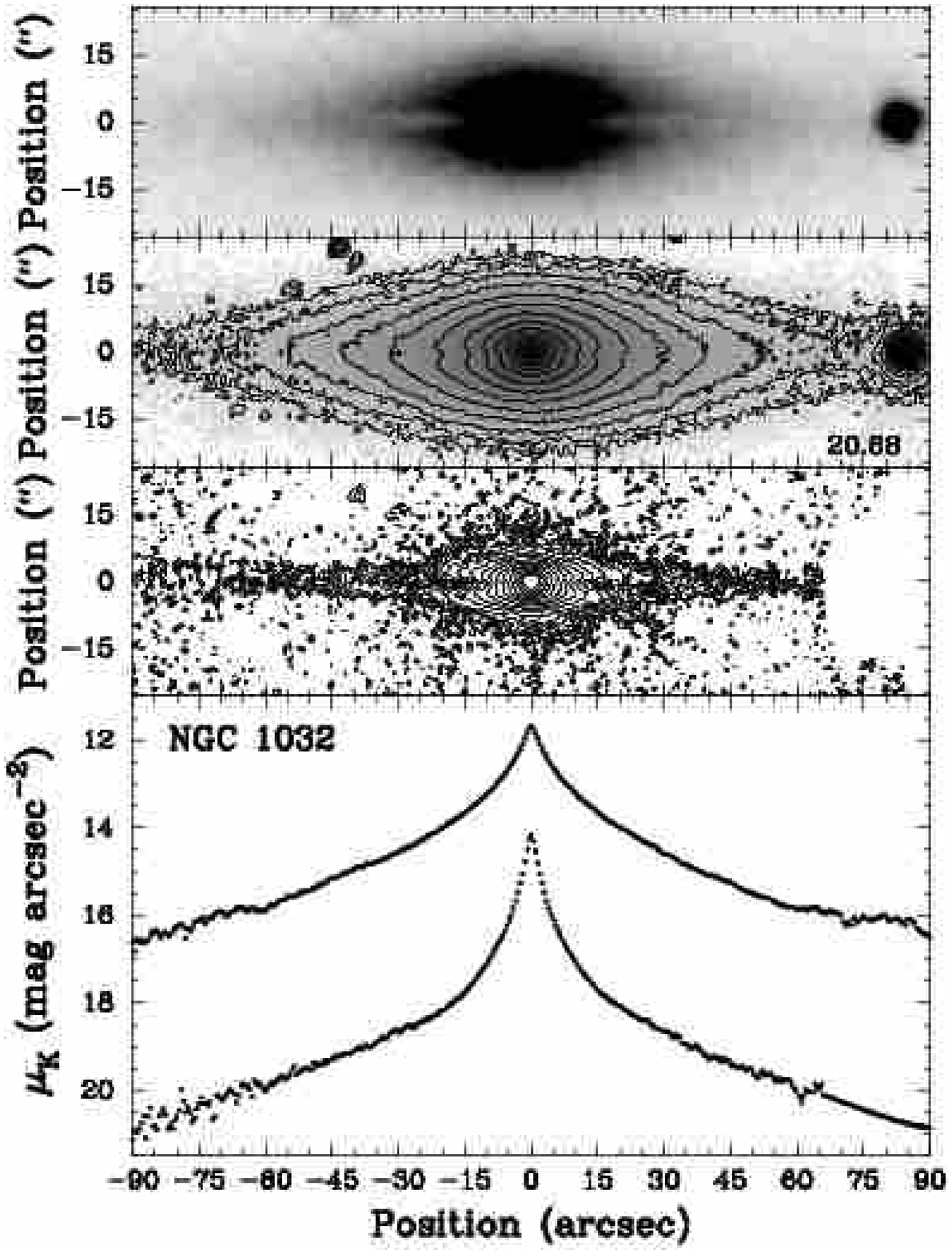}\hspace*{0.05\textwidth}\includegraphics[width=0.45\textwidth]{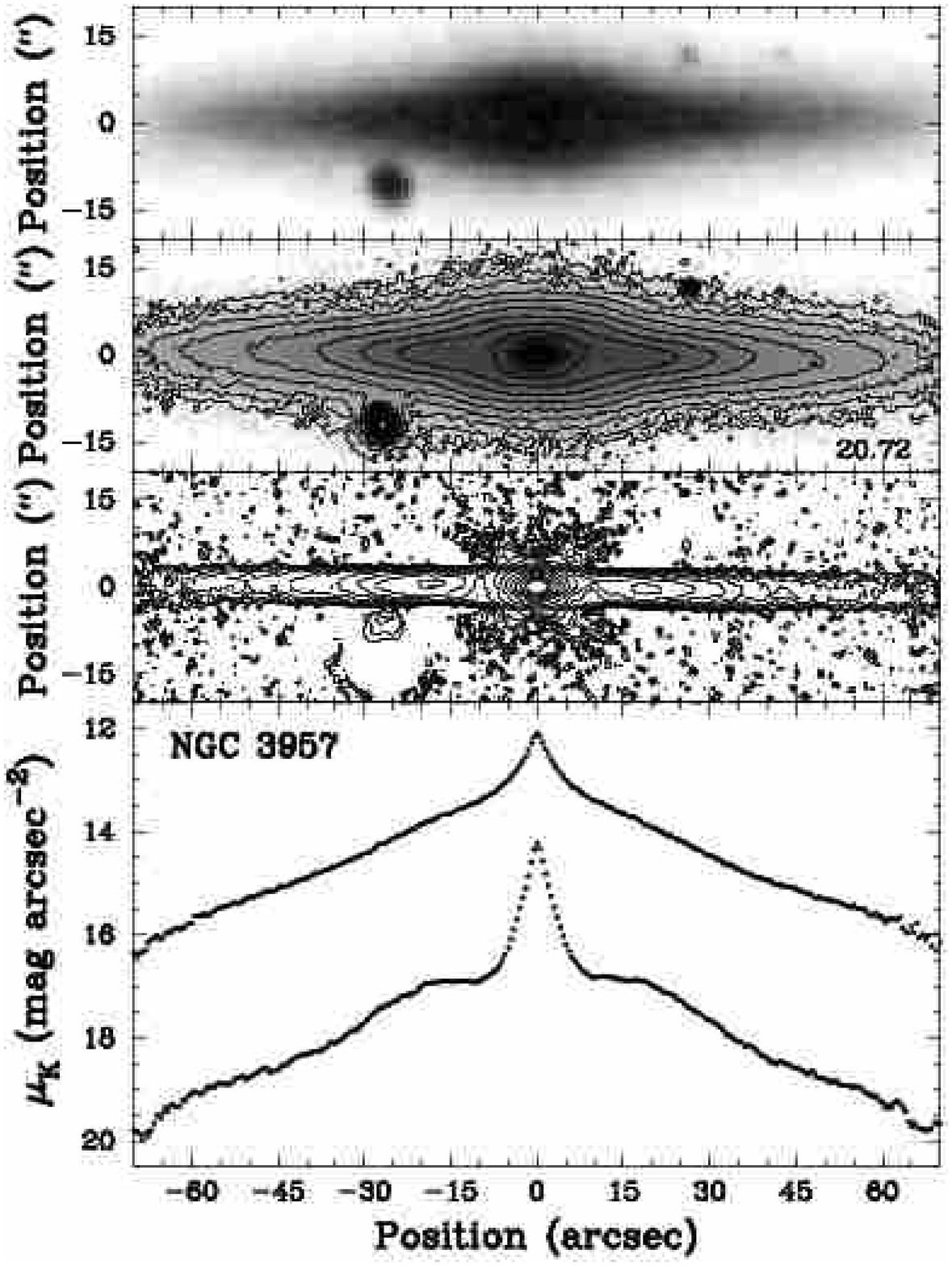}
\includegraphics[width=0.45\textwidth]{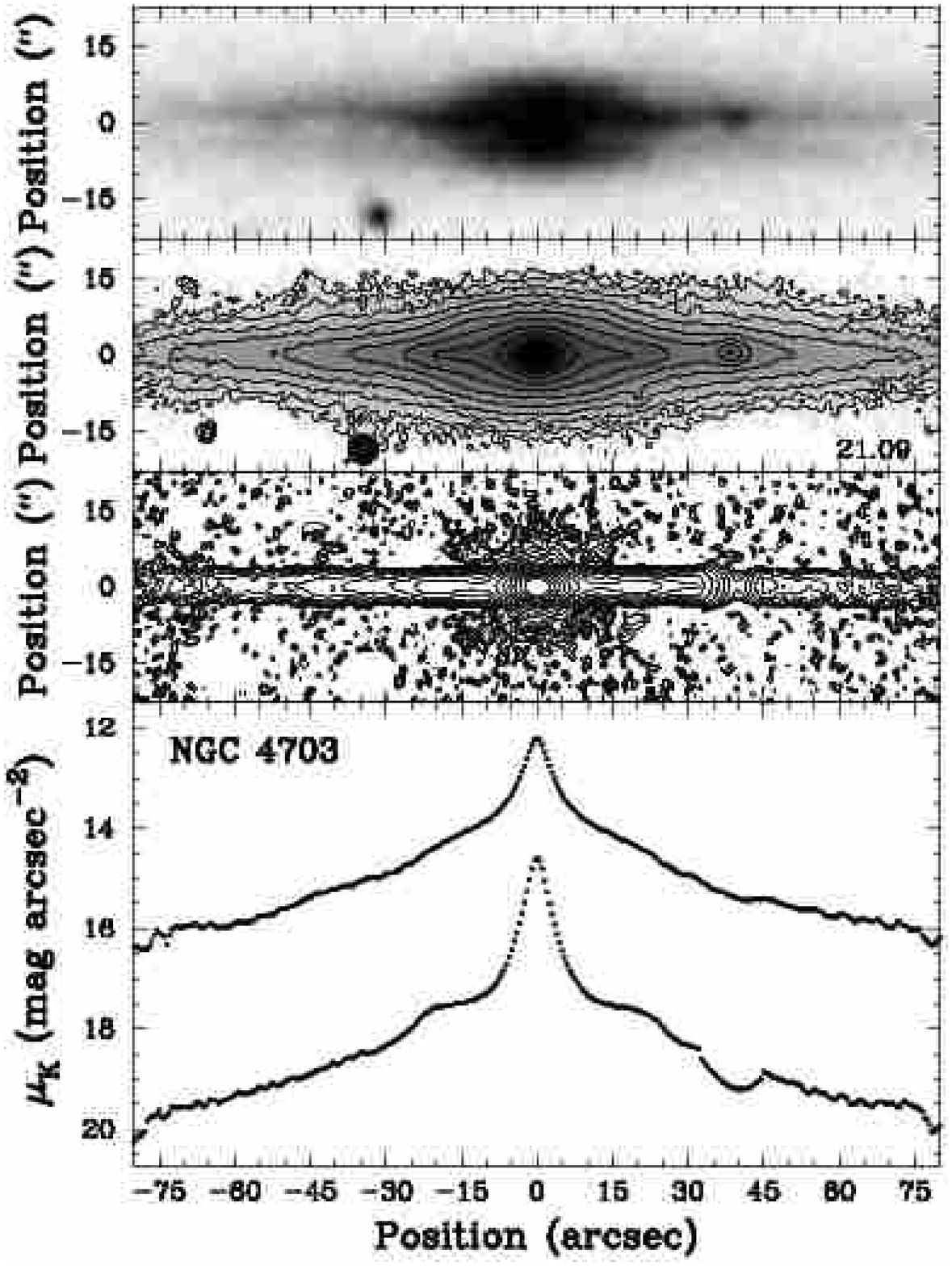}\hspace*{0.05\textwidth}\includegraphics[width=0.45\textwidth]{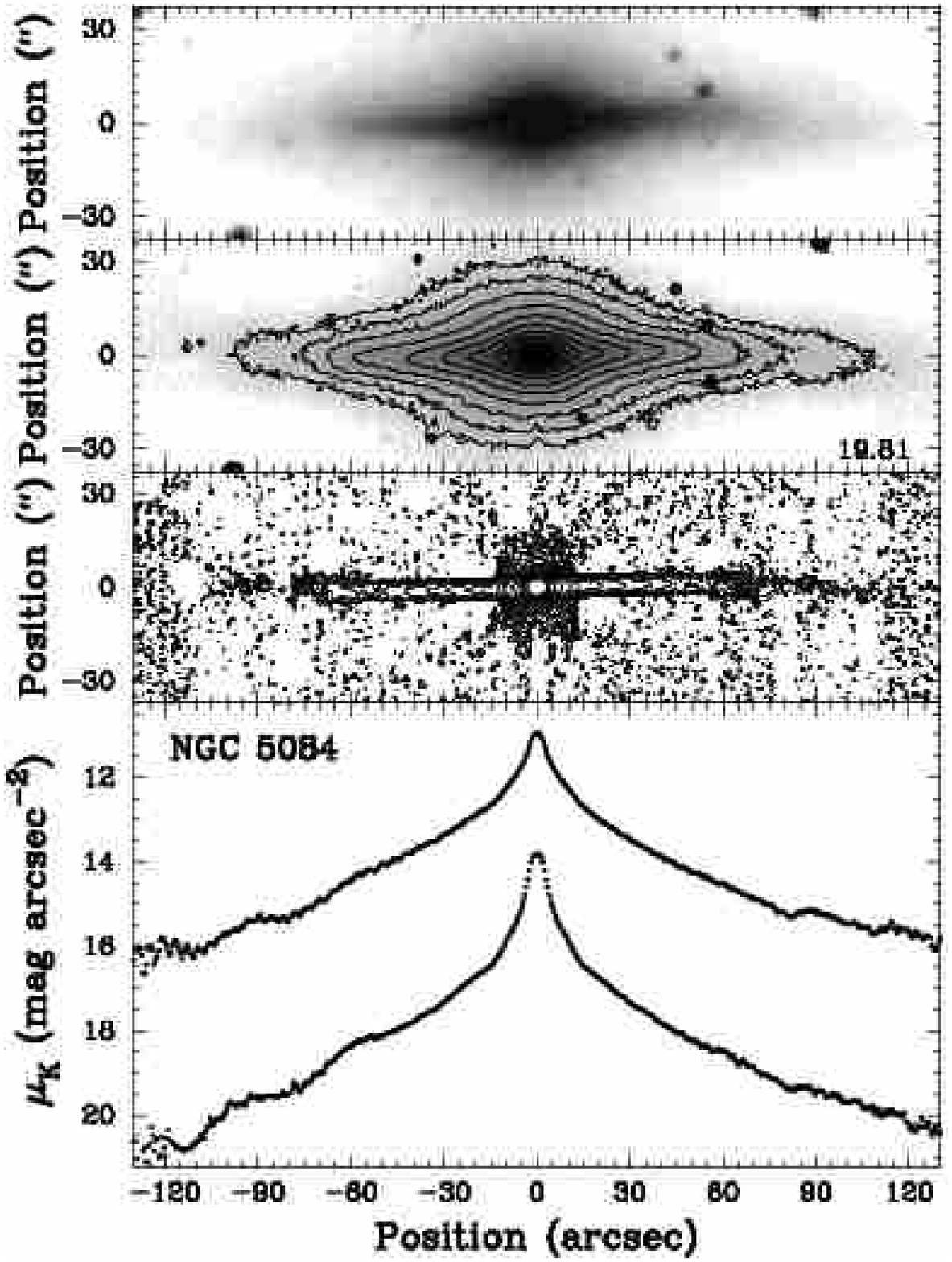}
\end{center}
\caption{Same as Figure~\ref{fig:bps} but for the control sample galaxies.}
\label{fig:control}
\end{figure*}
\begin{figure*}
\begin{center}
\includegraphics[width=0.45\textwidth]{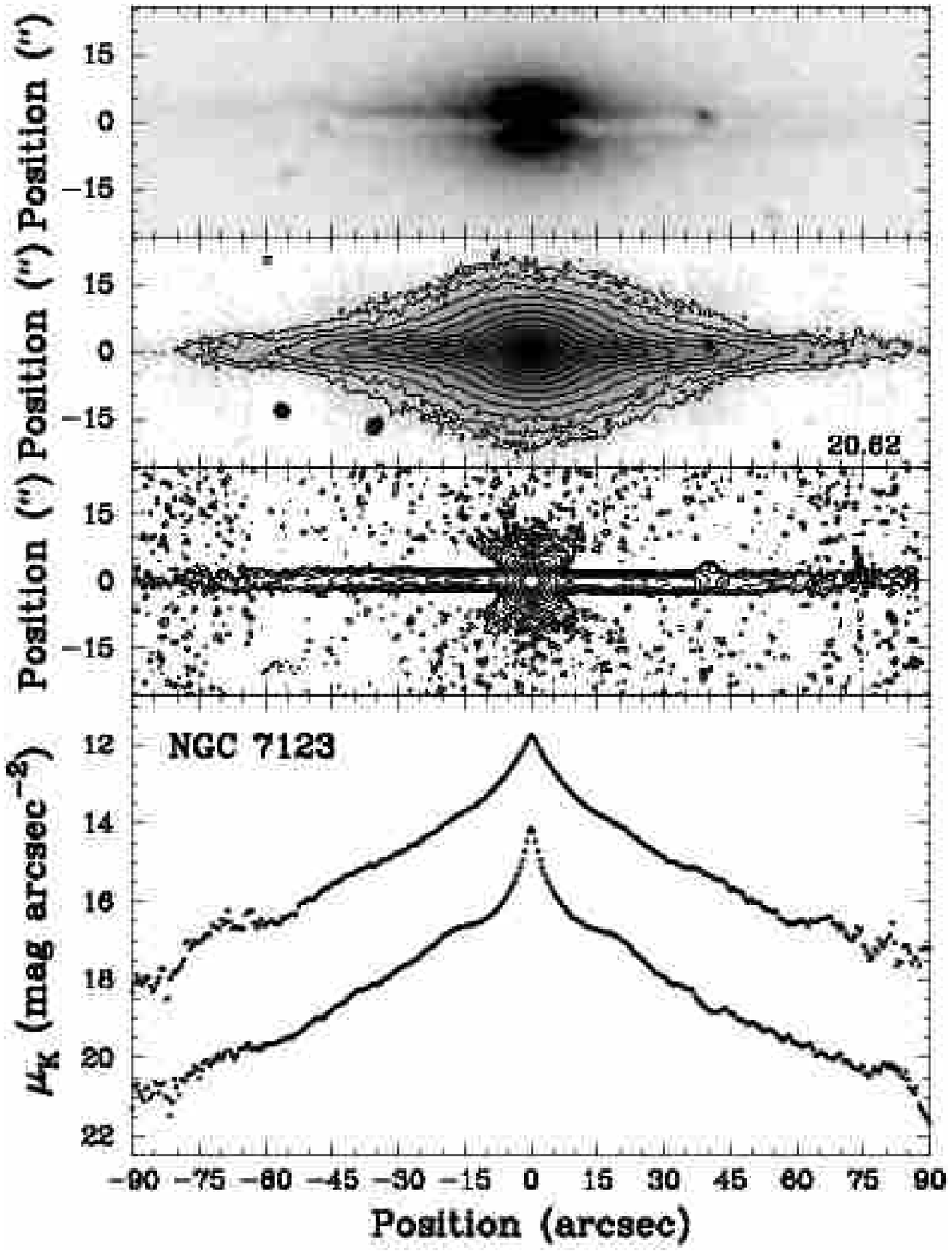}\hspace*{0.05\textwidth}\includegraphics[width=0.45\textwidth]{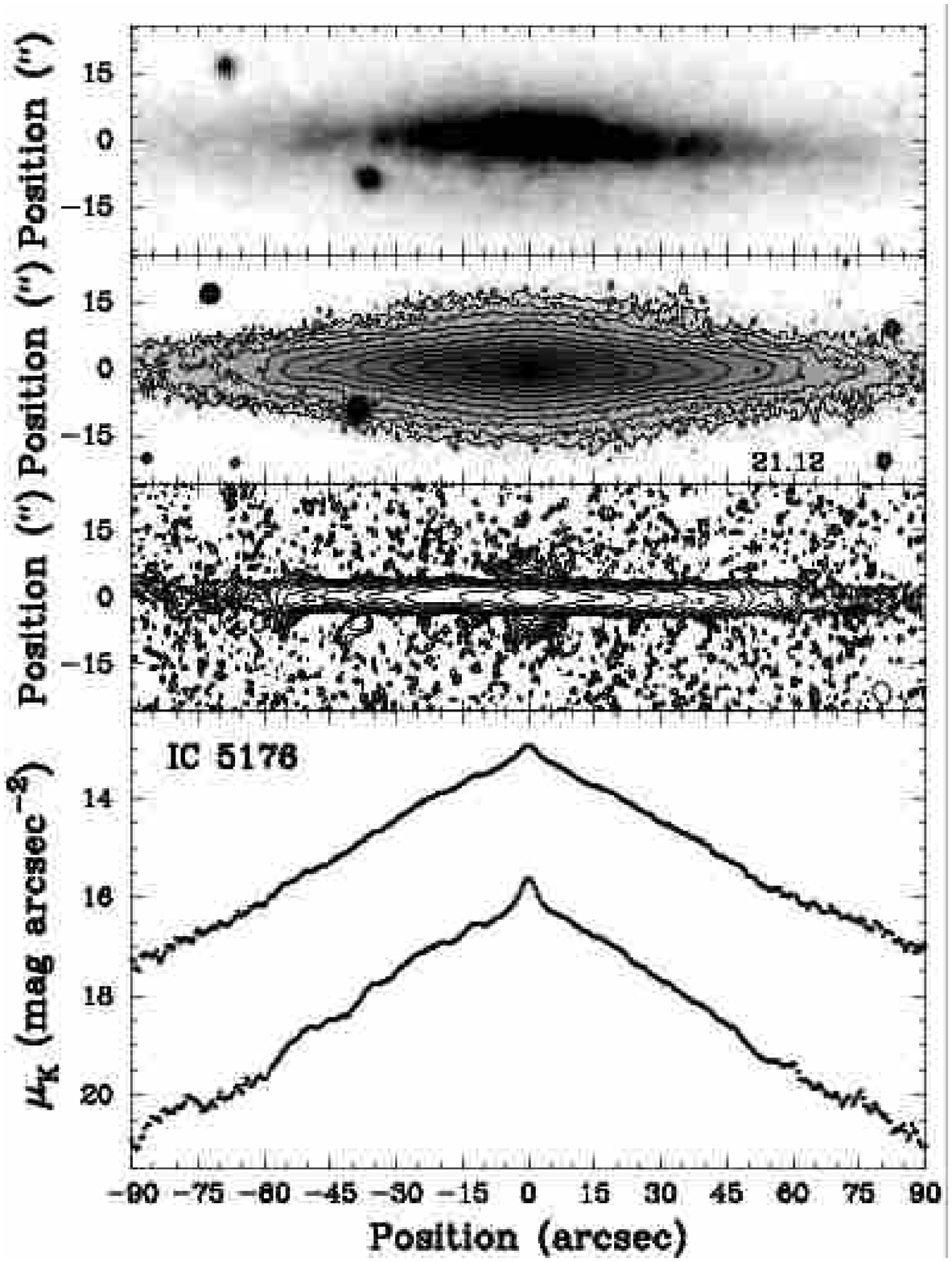}
\end{center}
\addtocounter{figure}{-1}
\caption{Continued.}
\end{figure*}

Comparing the DSS and \kn-band images, the advantages of working in
the NIR become obvious. Most importantly, dust extinction within the
galaxies, when present, is drastically reduced (see, e.g., IC~2531 and
NGC~7123), although it remains non-negligible in a few objects (e.g.\
NGC~1032). Generally, the B/PS and associated features are also
sharper at \kn-band, although this is hard to quantify. For a similar
reason, unsharp-masking, which enhances local extrema, is ideal to
highlight readily apparent morphological features and reveal weaker
ones.

The unsharp-masked images of Figures~\ref{fig:bps}--\ref{fig:control}
were obtained by median-filtering the \kn-band images, that is by
replacing the value of each pixel by the difference between it and
that of the median within a centered circular aperture. The size of
the aperture was held fixed across each image, but it was varied from
galaxy to galaxy to best highlight the features of interest here: X
shapes, secondary disc enhancements, \ldots\ Bigger apertures yield
unsharp-masked images more similar to the original \kn-band image,
while smaller apertures enhance smaller scale features such as central
discs, spiral arms and remaining dust lanes. The apertures used in
Figures~\ref{fig:bps}--\ref{fig:control} are thus tightly correlated
to the characteristic size of the morphological features enhanced. We
also note that the unsharp-masked images were obtained from \kn-band
images with the foreground stars interpolated over. This process is
not perfect, however, and can create artefacts in the unsharp-masked
images, especially near the equatorial plane of the galaxies where the
luminosity gradients are steep. One must thus be careful when
interpreting the results (see, e.g., NGC~2788A and NGC~6722).

As the same morphological features are present in many galaxies, we
provide a generic description of those features below and list the
specific features observed in each object in columns $2$--$7$ of
Table~\ref{tab:features}. We note however that, for some objects, some
features are best seen in images unsharp-masked on a spatial scale
different from that shown in
Figures~\ref{fig:bps}--\ref{fig:control}. To create
Table~\ref{tab:features}, we thus analysed images unsharp-masked with
circular apertures of radii ranging from $2.5$~arcsec ($5$ pixels) to
$17.5$~arcsec ($35$ pixels).

{\bf Off-centered X (OX).} Many galaxies show an X shape structure in
the bulge region, usually most easily visible in the unsharp-masked
\kn-band image. In many cases, however, if one follows the branches of
the X, the ridges do not cross in the centre of the galaxy but rather
fall short of it. This leads to a `$>\,<$' feature rather than a true
`$>\!<$', which we dub an off-centered X (OX). Excellent examples are
NGC~128 and NGC~6771, but there are many.

{\bf Centered X (CX).} In a comparable number of cases, the branches
of the X do cross in the center of the galaxy, and we dub this feature
a centered X (CX). Although the transition between off-centered and
centered X is somewhat ill-defined, good examples of centered X
features include NGC~1381 and IC~2531.

\begin{figure}
\begin{center}
\includegraphics[width=0.475\textwidth]{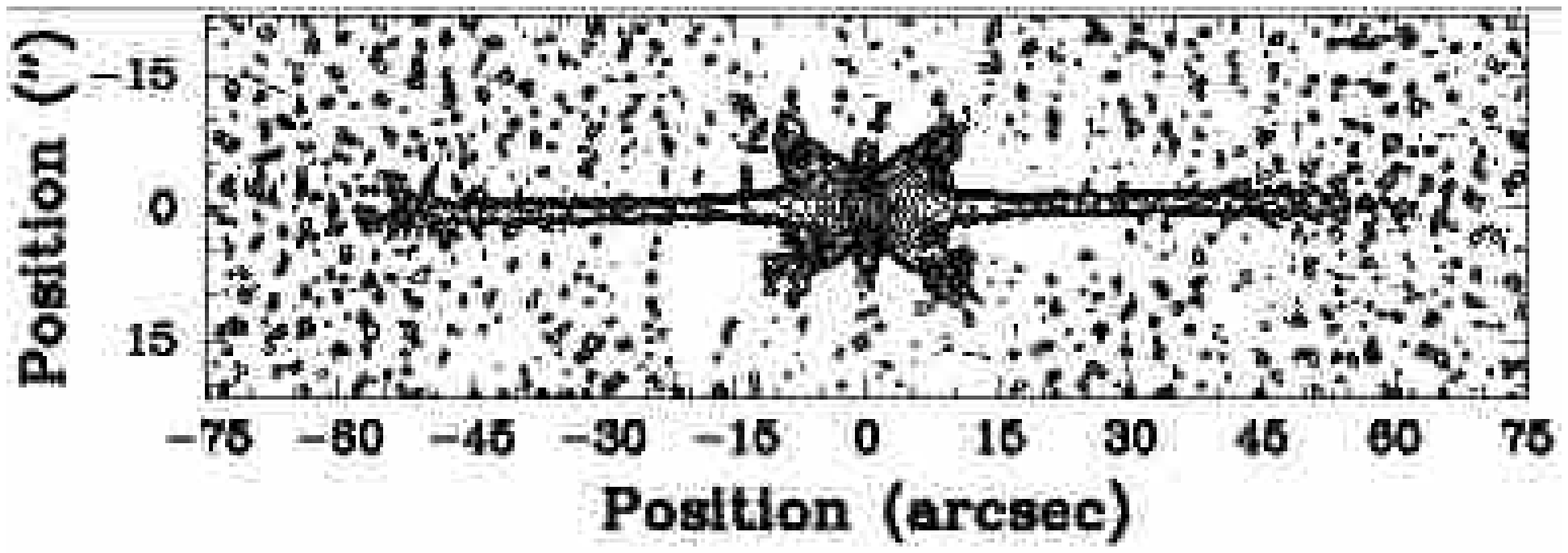}
\includegraphics[width=0.475\textwidth]{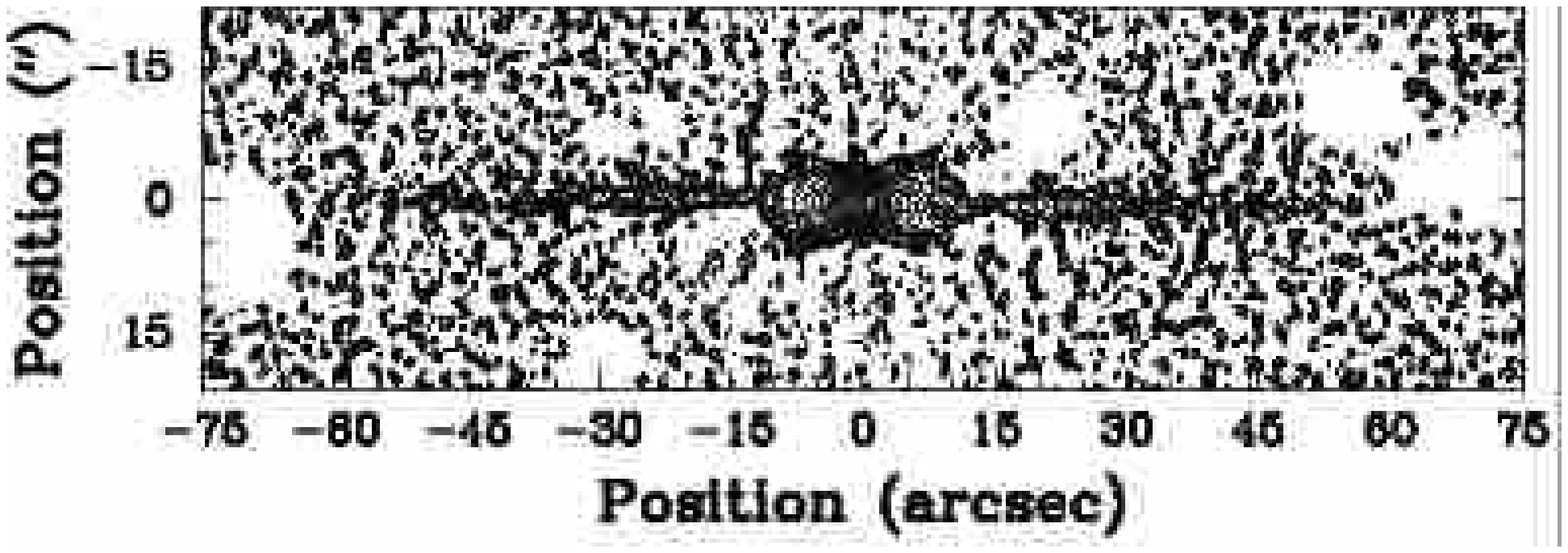}
\end{center}
\caption{\kn-band images of NGC~1381 (top) and NGC~3203 (bottom)
 unsharp-masked on a scale of $2.5$~arcsec. A long and narrow extremum
 is clearly seen along the minor-axis of NGC~1381, while a weaker one
 is observed in NGC~3203. We dub both features a minor-axis extremum
 (ME). NGC~3203 also reveals weak bisymmetric elongated maxima
 slightly offset from the major-axis, which we associate with spiral
 arms (SA) in a nearly but not exactly edge-on disc. Those spirals
 arms are not as apparent in images unsharp-masked on larger scales.}
\label{fig:me}
\end{figure}

{\bf Minor-axis extremum (ME).} Many galaxies show a rather narrow and
elongated local maximum along the minor-axis, which we dub minor-axis
extremum (ME). They are generally only visible in \kn-band images
unsharp-masked on a smaller scale than that presented in
Figures~\ref{fig:bps}--\ref{fig:control}, so we show the cases of
NGC~1381 and NGC~3203 in Figure~\ref{fig:me} for illustrative
purposes.
%
%

{\bf Secondary maxima (SM).} Many galaxies display local maxima away
from the center along the major-axis, which we dub secondary maxima
(SM). While sometimes readily visible in the original \kn-band image
(e.g.\ NGC~4710 and ESO~597-G036), most are easily seen only in the
unsharp-masked images (e.g.\ ESO~151-G004 and NGC~3203).

{\bf Spiral arms (SA).} A number of galaxies display narrow and
elongated local maxima which are not aligned with either axes but
rather are typically slightly offset from the major-axis, and
sometimes even partially wind around the bulge forming a
pseudo-ring. Those features are generally bisymmetric and are again
most easily visible in the unsharp-masked \kn-band images. We dub them
spiral arms (SA) for obvious reasons. Good examples include NGC~4469
and NGC~5746 (but see also the much weaker example of NGC~3203 in
Fig.~\ref{fig:me}).

In Table~\ref{tab:features}, we finally give an indication of the
inclination of the galaxies, labeling them either inclined (I), when
the galaxy is obviously more than a few degrees from exactly edge-on,
quasi edge-on (Q), when the galaxy is apparently away from edge-on by
at most a few degrees, and exactly edge-on (E), when there is no
indication at all that the galaxy is not exactly edge-on. There is
obviously a continuum of inclinations, but the first class generally
includes galaxies where spirals arms are easily visible (e.g.\
NGC~4469 and NGC~5746), while the second includes mostly galaxies with
weak spirals arms or a slightly off-centered (in the vertical
direction) dust lane (e.g.\ NGC~3203 and NGC~2788A). This
classification will be especially important in \citeauthor{aab06},
dealing with the vertical surface brightness profiles, since disc
projection effects can not be neglected for discs significantly away
from edge-on.
%
%
\begin{table*}
\begin{minipage}{177mm}
\caption{Galaxy features.}
\label{tab:features}
\begin{tabular}{llllllclll}
\hline
Galaxy & \multicolumn{6}{c}{Images} & \multicolumn{2}{c}{Profiles} & Notes\\
 & \multicolumn{6}{c}{$\overbrace{\hspace*{55mm}}$} & \multicolumn{2}{c}{$\overbrace{\hspace*{16mm}}$} & \\
 & OX & CX & ME & SM & SA & I & MA & SU & \\
\multicolumn{1}{c}{(1)} & \multicolumn{1}{c}{(2)} & \multicolumn{1}{c}{(3)} &
  \multicolumn{1}{c}{(4)} & \multicolumn{1}{c}{(5)} & \multicolumn{1}{c}{(6)}
  & \multicolumn{1}{c}{(7)} & \multicolumn{1}{c}{(8)}& \multicolumn{1}{c}{(9)} & 
 \multicolumn{1}{c}{(10)}\\
\hline
\noalign{\vspace{0.15cm}}
\multicolumn{8}{l}{\bf B/PS bulges} \\
NGC 128      & OX  &     & ME  & SM  &     & E & 3  & 3  & Warped disc ($m=0$ mode?)\\ 
ESO 151-G004 & OX  &     &     & SM  & SA  & Q & 3F & 3  & \\
NGC 1381     &     & CX  & ME  &     &     & E & 3  & 3  & Upwardly curved CX ($_\cap\!\!\!\!^\cup$)?\\
NGC 1596     & OX? &     & ME  &     &     & E & 2  & 2  & \\
NGC 1886     &     & CX  & ME  & SM  &     & Q & 3F & 3  & \\
NGC 2310     &     & CX  &     & SM  & SA  & I & 3F & 2  & Inner ring with outer spiral arms\\
ESO 311-G012 & OX  &     & ME  & SM  &     & E & 3  & 2  & \\
NGC 2788A    &     & CX  & ME? & SM  &     & Q & 4F & 4  & \\
IC  2531     &     & CX  &     & SM  &     & E & 3F & 2  & \\
NGC 3203     & OX  &     & ME  & SM  & SA  & Q & 3F & 2  & \\
NGC 3390     &     & CX  &     & SM  & SA  & Q & 3F & 2  & \\
NGC 4469     & OX  &     & ME? & SM  & SA  & I & 3F & 3  & \\
NGC 4710     & OX  &     &     & SM  & SA? & Q & 4F & 3F & Inner surface brightness plateau?\\
PGC 44931    &     & CX  & ME? & SM  & SA  & I & 3F & 2  & Inner ring with outer spiral arms?\\
ESO 443-G042 & OX  &     &     & SM  &     & Q & 3F & 3  & \\ 
NGC 5746     &     & CX  &     & SM  & SA  & I & 3F & 3  & Inner ring with outer spiral arms\\
IC  4767     & OX  &     &     & SM  &     & Q & 3F & 3  & \\
NGC 6722     & OX  &     & ME? & SM  & SA  & I & 3F & 2  & Inner ring with outer spiral arms, warped disc?\\
NGC 6771     & OX  &     & ME  & SM  &     & Q & 3F & 3  & \\
ESO 185-G053 &     & CX? &     & SM  &     & E & 3F & 3  & `( )'-shaped structure\\
IC  4937     & OX  &     &     & SM  & SA? & Q & 3F & 2  & \\
ESO 597-G036 & OX  &     & ME? & SM  &     & E & 3F & 3  & \\
IC  5096     &     & CX  & ME  & SM  &     & Q & 3  & 3  & \\
ESO 240-G011 &     & CX? & ME? &     & SA  & I & 3  & 2  & Inner ring with outer spiral arms\\
\hline
\noalign{\vspace{0.15cm}}
\multicolumn{8}{l}{\bf Control sample}\\
NGC 1032     &     &     & ME  &     &     & E & 2  & 2  & \\
NGC 3957     &     & CX  & ME  & SM  &     & Q & 3F & 2  & \\
NGC 4703     &     &     & ME  & SM  &     & Q & 4F & 3  & \\
NGC 5084     &     &     & ME  &     & SA? & Q & 2  & 2  & \\
NGC 7123     &     & CX  & ME  &     &     & E & 3  & 2  & \\
IC  5176     &     &     & ME  &     & SA? & E & 2  & 2  & \\
\hline
\end{tabular}
\\
Notes: For the images, the features listed were identified on images
unsharp-masked on scales ranging from $2.5$ to $17.5$~arcsec. OX:
off-centered X feature, CX: centered X feature, ME: minor-axis
extremum, SM: secondary maxima along major-axis, SA: spiral arms, I:
inclination. For the inclination, the galaxies are classified as
edge-on (E), quasi-edge-on (Q), or inclined (I). For the surface
brightness profiles, we list the number of clearly distinct
regions. MA: major-axis profile, SU: vertically-summed profile. The
letter `F' denotes a flat intermediate region.
\end{minipage}
\end{table*}
\subsection{Incidence of morphological features}
\label{sec:morph_incidence}
The most striking result from Figures~\ref{fig:bps}--\ref{fig:control}
is that the morphology of the galaxies with a B/PS bulge is much more
complex than that of the control sample galaxies, in the sense that
galaxies with a B/PS bulge contain more of the morphological features
described in \S~\ref{sec:morphology} (and best-revealed by
unsharp-masking; see Table~\ref{tab:features}). Indeed, at least $50$
per cent of galaxies with a B/PS bulge possess an off-centered X
feature and at least $38$ per cent a centered one, while only $33$ per
cent of the control galaxies have either. Similarly, $88$ per cent of
the galaxies with a B/PS bulge have secondary maxima along the
major-axis, while only $33$ per cent of the control galaxies do. For
the presence of spiral arms, the ratios become, respectively, at least
$38$ per cent and perhaps nil. Only for the minor-axis extrema are
those percentages inverted, with at most $58$ per cent of galaxies
with a B/PS bulge possessing one but all control galaxies. Those
statistics thus suggest that both centered and off-centered X
features, secondary major-axis maxima and spiral arms are
preferentially associated with B/PS bulges, while only minor-axis
extrema are found preferentially in other bulge types. In fact, the
contrast between the main and control samples would be even greater
if, as we will argue in \S~\ref{sec:disc_control}, the control sample
was not contaminated by weak B/PS bulges (NGC~3957 and NGC~4703).

We also note that, given the large number of secondary maxima observed
at $K$-band (in fact, more than in the optical images), it is unlikely
that they are due to the presence of obscuring dust filaments inclined
with respect to the equatorial plane, as was suggested by various
authors \citep[e.g.][]{s61,wb88}.
\subsection{Origin of morphological features and the orbital structure
of bars}
\label{sec:morph_origin}
Although the accretion of external material can give rise to X-shaped
features \citep[e.g.][]{bp85,hq88,hq89,wb88}, those are generally
centered and it is unlikely that any long-lasting off-centered X could
be produced. Since we observe a majority of off-centered X in our
sample, and since there is no obvious discontinuity between the two,
it is more attractive to look for a unique mechanism which can
simultaneously explain both centered and off-centered stable X
features.

\citet{a05} unsharp-masked a high-quality $N$-body simulation of a
barred disc. The results provide an essentially perfect match to the
secondary maxima discussed above and, depending on the viewing angle
to the bar, to the off-centered and centered X features, suggesting a
link between those features and edge-on bars. We discuss below the
possible origin of this link through the orbital structure of barred
discs and then discuss some caveats.

The most important orbit families in 3D bar models are those of the
$x_1$ tree, all elongated parallel to the bar and located within
corotation. They include the $x_1$ family itself (restricted to the
equatorial plane) and many other families bifurcating from the $x_1$
at vertical resonances of increasing energies. \citet{sw93} provide a
general introduction, but we adopt here the notation of
\citet{spa02a,spa02b} and call them $x_1v_1$, $x_1v_2$, $x_1v_3$,
etc. The morphological features identified in \S~\ref{sec:morphology}
can then be reproduced by superposing orbits of the appropriate
shapes, as done by \citet{psa02}. \citet{px06} recently explored very
similar ideas, with identical conclusions.

The centered and off-centered X features, in particular, can both
arise out of families extending out of the equatorial plane in the
vertical direction. The $x_1v_1$ family (also called banana orbits;
e.g.\ \citealt{pf91}), bifurcating from the $x_1$ family at the 2:1
vertical resonance, has the largest vertical extent and is shaped like
a smile or a frown. As shown by the orbit superpositions of
\citet{psa02}, $x_1v_1$ orbits seen side-on have a morphology entirely
consistent with an off-centered X. Depending on the model (i.e.\ mass
distribution and pattern speed) and viewing angle, centered X features
can also be created.

The $x_1v_3$ and $x_1v_4$ orbit families (bifurcating from the $x_1$
family at the 3:1 vertical resonance and shaped like a `$\sim$') and
the $x_1v_5$ orbit family (bifurcating at the 4:1 resonance and shaped
like a `w') can also give rise to centered or off-centered X features,
depending on the model and viewing angle \citep[see,
again,][]{psa02}. This is also the case for the $z_{3.1}s$ family,
which does not bifurcate from the $x_1$ orbits but is morphologically
similar to the $x_1v_4$ family, although it was present in only one
model.

The orbit superpositions of \citet{psa02} further show that the orbit
families described above (as well as higher order families such as
$x_1v_7$, $x_1v_8$ and $x_1v_9$) can give rise to a number of maxima
along the major-axis, similar to the secondary maxima identified in
\S~\ref{sec:morphology}. Those maxima generally occur at larger radii
than the X features and near (but within) the ends of the bar, often
where orbits have loops. An alternative and more straightforward
explanation is that the secondary maxima observed are simply the
edge-on projections of inner rings (edge brightening), known to exist
in a large fraction of barred spiral galaxies and located at the end
of the bar \citep[e.g.][]{k79,b95}. Inner rings naturally form in
gas-rich discs under the influence of rotating bars
\citep[e.g.][]{s81,s84,brsbc94}, but they also develop in purely
dissipationless simulations \citep[e.g.][]{am02,ba05}. Either way,
both mechanisms support a relationship to bars, and both mechanisms
may coexist. Accretion scenarios producing centered X features, on the
other hand, will generally not produce secondary maxima along the
major-axis (even less so systematically located outside of the X
feature itsef), making them even less attractive.

The orbit superpositions of \citet{psa02} show that the barred orbit
families considered above can also give rise to the minor-axis extrema
described in \S~\ref{sec:morphology}, especially when seen end-on. The
$z_i$ orbits are particularly interesting in this respect but were
little studied by \citet{spa02a,spa02b} and deserve more
attention. This is encouraging, but the fact that all control sample
galaxies also show a minor-axis extremum suggests that this feature is
not related to B/PS bulges (and thus presumably to bars). We do not
properly understand the origin of this feature.

As we mentioned in \S~\ref{sec:morphology}, because of the
morphological similarities, we interpret the local maxima elongated
parallel to but slightly offset from the major-axis of the galaxies as
spiral arms. If this association is true, which we have no reason to
doubt, then obviously spiral arms can only be detected in galaxies
that are not perfectly edge-on. In fact, we have used the presence of
spirals to quantify the inclination of the galaxies in
Table~\ref{tab:features}. Since we have no other independent and
reliable way to estimate the inclination of the galaxies precisely (as
we do not know the intrinsic thickness of the discs), we can not
guarantee that the inclination distribution is similar between the
sample of galaxies with a B/PS bulge and the control sample. Thus, a
reliable comparison of the incidence of spiral arms between the two
samples is not possible, and we do not discuss this feature at length.

Suffice it to say that the apparently slightly higher incidence of
spiral arms in galaxies with a B/PS bulge is consistent with the
suggestion that B/PS bulges are simply thick bars viewed
edge-on. Indeed, bars are very good at driving (grand design) spiral
patterns \citep[e.g.][]{a80,tw84,ee89} and spiral arms are
preferentially found in barred galaxies
\citep[e.g.][]{ee89,bvsl05}. Furthermore, for the few galaxies where a
(partial) inner ring with outer spiral arms is visible (e.g.\
NGC~5746), the ring is systematically located just at the end of the
flat intermediate region of the major-axis surface brightness profiles
(see \S~\ref{sec:sbprofs}; all such galaxies have an intermediate
region, and it is flat in all cases but one). If, as we argue in
\S~\ref{sec:sbprofs}, the end of the intermediate surface brightness
profile region also marks the end of the bar, then this trend is
entirely consistent with observations of more face-on barred spiral
galaxies \citep[e.g.][]{k79,b95} and the expectations from theory
\citep[e.g.][]{s81,s84,brsbc94}.

On a different note, we must point out that while the (barred) picture
proposed above is attractive and supported by much independent
evidence (see \S~\ref{sec:intro}), it may not be unique. Indeed, as
simple tests show, surface brightness distributions created from
nested rectangular isophotes can also give rise to centered and
off-centered X-shapes when unsharp-masked, independently of their
origin. As axisymmetric boxy distributions can in principle also exist
\citep*[e.g.][]{bp85,man85,r88}, it is possible that bars are not
uniquely related to the complex features observed. However, as the
N-body simulations and orbit properties discussed above show, barred
models are consistent with the unsharp-masked features in much
details, while this remains to be shown for other models.
%
%
\section{SURFACE BRIGHTNESS PROFILES}
\label{sec:sbprofs}
\subsection{Profile features}
\label{sec:sbp_features}
In addition to the DSS, \kn-band and unsharp-masked \kn-band images,
the panels of Figures~\ref{fig:bps}--\ref{fig:control} also contain
two surface brightness profiles. The fainter ones represent the
major-axis surface brightness profiles, extracted using a constant
position angle, an approximation that is valid in the inner parts of
all galaxies and in the outer parts of all but a few (e.g.\ NGC~128,
NGC~6722, NGC~6771 and ESO~597-G036). The brighter surface brightness
profiles were obtained by summing the data in the vertical direction,
as if the galaxies were infinitely thin, again assuming a constant
position angle for the major-axis. In both cases, the \kn-band images
with the foreground sources interpolated over were used, and the
vertical profiles were summed until the noise level of the image was
reached. While the noise level varies slightly from object to object,
such a limit is sufficient for our purposes.

While our summed surface brightness profiles are not strictly
equivalent to azimuthally-averaged surface brightness profiles in more
face-on galaxies, or to cuts at a given position angle (i.e.\ they are
also integrated along the line-of-sight through the edge-on discs), a
comparison is nevertheless possible and useful. Indeed, because the
luminosity of discs generally decreases rapidly with radius, the light
at any given position in our profiles is still dominated by that near
the same (cylindrical) radius. Our profiles can also be compared to
those in \citet{ldp00b}, who showed profiles (cuts) parallel to but
offset from the major-axis for a number of edge-on spiral galaxies.

\begin{figure}
\begin{center}
\includegraphics[width=0.475\textwidth]{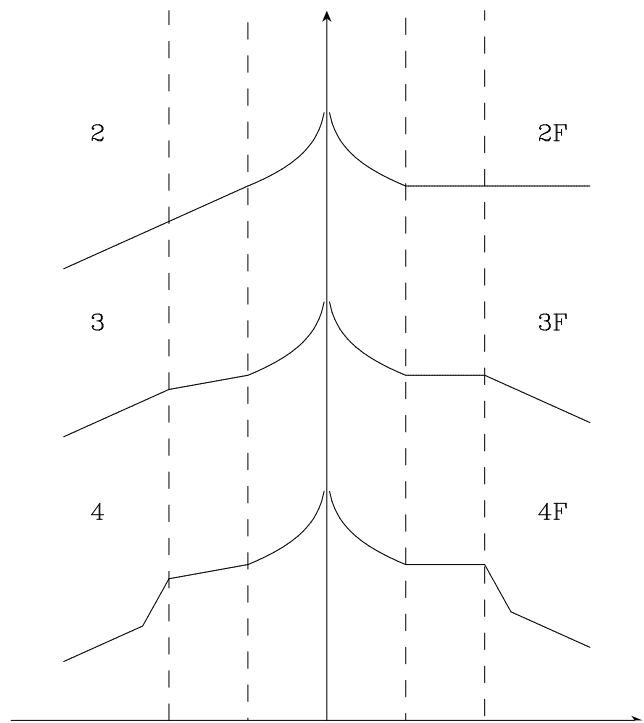}
\end{center}
\caption{Cartoon description of our surface brightness profiles, as
  tabulated in Table~\ref{tab:features}. Two, three, and four-region
  surface brightness profiles are shown, with or without a flat
  intermediate region (denoted by the letter `F'). Note that no `2F'
  profile was actually observed.}
\label{fig:sbprof_cartoon}
\end{figure}

In columns $8$ and $9$, Table~\ref{tab:features} lists the number of
distinct regions present in the surface brightness profiles of each
galaxy, each region being typically separated by a clear radial break
(although one must be cautious with badly interpolated foreground
sources). From the surface brightness profiles of face-on galaxies, we
would generally expect each profile to show only two distinct regions
if the galaxy is axisymmetric. A first region at small radii,
generally associated with the bulge, where the profile is steep
(generally steeper than an exponential profile), and a second region
at larger radii, generally associated with the disc, where the profile
is exponential\footnote{We remind the reader that the surface
brightness profile of an edge-on exponential disc is not exactly
linear in a (logarithmic) surface brightness vs.\ projected radius
plot, but slightly curved \citep[see, e.g.,][]{pdls00}.}. However,
such two-region profiles are rare in our data, especially along the
major-axis (but see NGC~1596 and NGC~1032). Most profiles show an
additional region at intermediate radii, where the profile is
shallower than the outer disc, even often flat or slightly rising with
radius (denoted by the letter `F' in Table~\ref{tab:features}; e.g.\
ESO~151-G004 and NGC~2310). Such three-region surface brightness
profiles are often referred to as Freeman Type~II profiles in the
literature \citep{f70}. A few galaxies also show two distinct
intermediate regions, one shallow following the steep central
component and one steep preceding the outer exponential disc, for a
total of four regions (e.g.\ NGC~2788A and NGC4710). We do not discuss
(or count) here apparent radial breaks in the outer discs, since they
are presumably only weakly related to the bulge
structure. Figure~\ref{fig:sbprof_cartoon} shows a cartoon version of
the various types of profiles identified in Table~\ref{tab:features}.
%
%

\subsection{Incidence of profile features}
\label{sec:sbp_incidence}
Again, the most obvious and basic result from
Figures~\ref{fig:bps}--\ref{fig:control} is that the surface
brightness profiles of the galaxies with a B/PS bulge are much more
complex than those of the control sample galaxies, especially along
the major-axis, in the sense that the surface brightness profiles of
galaxies with a B/PS bulge typically contain a larger number of
distinct regions separated by clear radial breaks (see
Table~\ref{tab:features}). Indeed, along the major-axis, $96$ per cent
of the galaxies with a B/PS bulge have a surface brightness profile
with $3$ or more regions (Freeman Type II profile), $78$ per cent of
which have a flat intermediate region, while the corresponding
fractions for the control sample are, respectively, only $50$ and $66$
per cent. The statistics are not as sharply contrasted when the
profiles summed along the vertical axis are considered, but the
differences remain significant. For example, $58$ per cent of the
galaxies with a B/PS bulge have a summed surface brightness profile
with a least $3$ regions, while only $17$ per cent of the control
galaxies do. Those fractions are in rough agreement with those of
\citet{ldp00b}, who carried out a similar analysis. As for the
morphological features, it thus appears that complex surface
brightness profiles are preferentially associated with B/PS
bulges. And again, the contrast between the main and control samples
would be sharper if the control sample had been better selected (i.e.\
if NGC~3957 and NGC~4703 has been excluded; see
\S~\ref{sec:disc_control}).
\subsection{Origin of profile features and relationship to bars}
\label{sec:sbp_origin}
The fact that practically all galaxies with a B/PS bulge have a
major-axis surface brightness profile with an intermediate, often flat
region is particularly important, as it is inconsistent with an
axisymmetric bulge and exponential disc model, no matter how steep the
bulge profile is (i.e.\ no matter what the Sersic $n$ index of the
bulge is). There is thus apparently a third photometric or
morphological `component' dominating at intermediate radii. We argue
below, however, that in such Freeman Type~II profiles, both the steep
inner region {\em and} the flat intermediate region are ultimately
caused by a single bar viewed edge-on, without the necessity of a
classic bulge.

\citet{ba05} followed the temporal evolution of the major-axis surface
brightness profile in a large number of barred $N$-body simulations
viewed edge-on. They showed that bar formation and evolution within an
(initially) exponential disc is associated with the buildup and
continued growth of a dense central region, resulting in a central
peak (which would normally be identified with a bulge) and with the
formation and gradual flattening of an intermediate region, in
addition to the outer exponential disc (see also
\citealt{am02,a03}). The central peak extends to $1$--$1.5$ original
disc scalelengths while the intermediate region extends to the end of
the bar, well beyond the central peak and the thickest part of the
bar. The existence of three surface brightness profile regions is thus
totally normal (and expected) in barred galaxies, but it remains
unexplained in classic bulge models, unless there is an additional
bar. In fact, four-region surface brightness profiles are also often
produced in barred models (e.g.\ Fig.~3 in \citealt{ba05}).

As expected from the elongated boxy/peanut shape of the bar, the ratio
of the length of the thickest part of the bar (or that of the central
peak) to that of the flat intermediate region increases in $N$-body
simulations as the viewing angle increases (from end-on to
side-on). This trend is also present in our data, where the ratio of
the length of the thickest part of the bulges (or that of the central
peak) to that of the intermediate region is generally larger in
peanut-shaped bulges than in boxy ones (compare, e.g., NGC~5746 and
NGC~6722 to NGC~1381 and IC~5096). There is however much variety, as
many strong peanut-shaped bulges have a small ratio (see, e.g.,
NGC~1886 and ESO~443-G042), and likely many causes for it. First, as
far as the surface brightness profile and bar shape are concerned,
there is a certain degeneracy between bar strength and viewing angle
(a strong bar seen at intermediate viewing angle may appear similar to
a weaker bar seen exactly side-on; e.g.\ \citealt{ba05}). Second, the
ratio described evolves in $N$-body simulations, generally decreasing
with time as the bar strengthens and lengthens but the central
component changes little \citep[see, again,][]{ba05}. Third, gas
inflow and subsequent star formation (absent from the pure $N$-body
models) may substantially modify the light distribution in the central
regions \citep[e.g.][]{hs94,fb95}. Fourth, the presence (or absence)
of rings and lenses will complicate the comparison. We also note that
as the profiles of \citet{ba05} were integrated vertically over $0.5$
original disc scalelengths, they are somewhat intermediate between our
major-axis and summed profiles.

In fact, the variety observed in the ratios of the length of the
thickest part of the bulge (or that of the central peak) to that of
the intermediate region may well be dominated by the range of bar
strengths present in the sample, rather than by the range of viewing
angles. This is also suggested by the data of \citet{ldp00b}, and it
would similarly imply that the observed boxy-peanut `sequence' is not
mainly one of viewing angle, but is instead driven by a varying (i.e.\
increasing) bar strength. The small ratios observed in strong
peanut-shaped bulges such as NGC~1886 and ESO~443-G042 can then be
explained only if the central surface brightness peak and the thick
part of the bulges are shorter in stronger bars (in a relative
sense). This is natural if the inner part of the surface brightness
profiles is dominated by a disc-like bulge, as will be argued in
\S~\ref{sec:disc_3d}. Indeed, \cite{a92a,a92b} showed that the size of
(bar-driven) nuclear gas discs is limited by the outer inner Lindblad
resonance (i.e.\ by the outermost $x_2$ orbit, elongated perpendicular
to the bar), which itself inversely depends on the bar strength. Then,
the stronger the bar, the smaller the nuclear disc.
%
%
\section{DISCUSSION}
\label{sec:discussion}
\subsection{Radial redistribution of material}
\label{sec:disc_radial}
\citet{a02,a03} argued that much of the bar-driven evolution in discs
is due to a transfer of angular momentum from the inner (barred) disc
to the outer disc (beyond corotation) and the dark halo, leading to a
strengthening, lengthening and slowing down of the bar with time. To
isolate the effects of angular momentum exchange, which leads to a
{\em radial} redistribution of matter, we must preferentially consider
the vertically-summed surface brightnes profiles of
Figures~\ref{fig:bps} and \ref{fig:control}. While fewer galaxies with
a B/PS bulge exhibit $3$ or more regions in the summed profile than
along the major-axis, a majority still do, while only one of the
control sample galaxies does (see Table~\ref{tab:features}).

In scenarios where the formation and evolution of discs is dominated
either by collapse or the gradual accretion of discrete components
(e.g.\ cannibalized dwarf or satellite galaxies), there is no obvious
mechanism to create the radial breaks in the disc density distribution
observed in our sample galaxies, even less to explain the spatial
correlation of those breaks with the ionized-gas and stellar
kinematics \citep[see][]{bf99,cb04}. Similarly, the traditional
interpretation of Freeman Type~II profiles as inwardly truncated discs
neither proposes an origin for the truncation nor explains why the
truncation is only partial and the surface brightness profiles
systematically have the specific shape observed. In bar-driven
evolution scenarios, however, a break naturally occurs at the end of
the bar \citep[e.g.][]{am02,a02,a03,ba05}. If this is indeed the case
here, then the break in a surface brightness profiles at the end of
the intermediate region should systematically mark the bar's end.

Our major-axis surface brightness profiles match those of \citet{am02}
and \citet{ba05}, and as we mentioned already the kinematics of the
sample galaxies agree very well with barred galaxy models
\citep[see][]{bf99,ab99,cb04,ba05}. A quick comparison further shows
that the position where the rotation curve flattens, generally
associated with the end of the bar \citep[e.g.][]{ba05}, always occurs
very near (although generally slightly within) the break in the
\kn-band surface brightness profile. Thus, our surface brightness
profiles are consistent with those expected from bar-driven evolution
scenarios, and the complex surface brightness profiles observed appear
to be a direct consequence of the transfer of angular momentum and the
radial (and vertical) rearrangement of material by the bar.

Most barred galaxies appear to have fast bars, whereby corotation,
defined as the radius where the stars and bar pattern rotate at the
same speed (and marking the transition between material loosing and
gaining angular momentum), is located just beyond the end of the bar
\citep*[see, e.g.,][]{gkm03,adc03}. Since the pattern speeds of the
bars in our sample are unknown, a direct test of this is
impossible. However, for the few galaxies with an apparent inner ring
(see Table~\ref{tab:features}), the ring radius is systematically
(roughly) equal to that of the break at the end of the intermediate
surface brightness region. Given that inner rings are generally
thought to occur near the inner 4:1 (ultra-harmonic) and corotation
resonances \citep*[e.g.][]{s81,s84,abcs82,b95,psa03b}, our galaxies
are consistent with harbouring fast bars.

It is also interesting to note that the two latest galaxies in our
sample, ESO~240-G011 and IC~5176 (with very small bulges and thus
presumably very weak bars, if any) both have completely featureless
(single) exponential outer discs. They thus offer useful benchmarks
against which more complex systems can be compared.
\subsection{Vertical redistribution of material}
\label{sec:disc_vertical}
To isolate the effects of the vertical redistribution of material
within bars, as predicted by models where the bar buckles and thickens
\citep[e.g.][]{cs81,cdfp90,rsjk91}, we compare here the major-axis and
vertically-summed surface brightness profiles of our sample galaxies.

A cursory examination of our major-axis and summed surface brightness
profiles shows that, for most galaxies with a B/PS bulge, they differ
markedly from one another. As already pointed out, the
vertically-summed profiles generally show fewer of the features
described in \S~\ref{sec:sbprofs} and they approach more closely a
simple bulge plus exponential disc description, while the major-axis
profiles generally show a steeper inner peak and a flatter
intermediate region (the outer regions have a comparable slope in both
profiles). Excellent examples of this are ESO151-G~004 and PGC~44931,
but there are many.

If the scaleheight of the stars in a galaxy is constant with radius,
one expects the major-axis and vertically-summed profiles to have the
same functional form but different zero-points, i.e.\ be offset but
parallel to one another (although not exactly, since the profiles are
vertically-summed to a given surface brightness level rather than to a
fixed number of scaleheights). This is not the case for most galaxies,
however, even when the major-axis surface brightness profile is flat
(when summing to a given surface brightness level {\em is} equivalent
to summing to a fixed number of scaleheights), suggesting that the
stellar scaleheights are not constant. See, again, ESO~151-G004 and
PGC~44931, although any galaxy with a flat intermediate profile will
do.

By discussing a single scaleheight, even in regions where the inner
peak of the surface brightness profiles is non-negligible, we have
however implicitely amalgamated the bulge and disc (no matter how the
former is defined). In the classic bulge and disc model, where the
bulge and disc represent structurally and kinematically distinct
compoments, this will necessarily lead to a variation of the
scaleheight, although one would naively expect this variation to be
monotonic. The major-axis and vertically-summed surface brightness
profiles of Figure~\ref{fig:bps}, however, show that the variations
are not monotonic. Furthermore, the functional difference between the
two profiles is systematically greatest in the flat intermediate
region, which would never be associated with a bulge (even in classic
models) and is clearly dominated by disc material. Our major-axis and
vertically-summed surface brightness profiles thus show that the
radial scaleheight variations are real and that they occur in the
discs (i.e.\ they are {\em not} due to bulge material), in direct
contradiction to the long accepted wisdom that disc scaleheights are
radially constant \citep[see,
e.g.,][]{ks81a,ks81b,ks82a,ks82b,sg90,gk96}. In fact, our results
support a picture where the bulge and disc are not intrinsically
distinct (either structurally or kinematically), but both instead
emerge from the rapid radial variation of the scaleheight of the disc
material (due to bar-related vertical resonances). This will be
discussed in more depth in \citeauthor{aab06}, where it will also be
shown that the variations are as expected from barred $N$-body
models. Viewed simplistically, our observations argue for the
reassignment of (most of) the bulge material to the disc (a B/PS
bulge), but this is an important conceptual change.

To our knowledge, a variable stellar scaleheight has only been
identified and discussed in a few galaxies. In the outer parts of
NGC~3115, \citet{cvh88} assign the variation to the end of the disc
self-gravity, while in the inner parts of the Milky Way \citet*{kdf91}
assign it to a variable ratio of young to old stars (their data in
fact already show strong hints of the peanut shape of the Galactic
bulge and a Freeman Type~II profile). \citet{gp97} do note an increase
of the scaleheight with (projected) radius in a large number of
edge-on spiral galaxies, but they assign the increase to the presence
of a thick disc with scalelength and scaleheight larger than those of
the dominant (thin) disc component. Upon closer inspection, a number
of other studies arguing for a constant scaleheight as a function of
(projected) radius actually do detect variations, but those are
generally monotonic and are argued to be consistent with the influence
of a structurally distinct bulge (in the center) or a thickening of
the outer disc (near the edge of the optical disc). The scaleheight
variations are thus rejected as artefacts unrelated to the dominant
disc, presumed thin \citep[e.g.][]{gk96}. While this is defendable in
some cases, it is doubtful in others, and the non-monotonic variations
observed here and in \citeauthor{aab06} in the inner and intermediate
regions of our sample galaxies can not be explained away in such
manners. Possible variations in the outer parts of our galaxies (e.g.\
flaring) are not of direct interest to the bulges' structure.
\subsection{Classic versus `pseudo'-bulges}
\label{sec:disc_bulges}
As discussed also in \citet{a05}, galaxy bulges have traditionally
been defined either as 1) the steep central component of the surface
brightness profile (generally steeper and brighter than the inward
extrapolation of the outer exponential disc; e.g.\ \citealt*{cfw99}),
2) the thick galactic component (in isophotal terms, clearly sticking
out of the equatorial plane in nearly edge-on objects) or 3) the
kinematically hot component (the central peak in the velocity
dispersion profile). Those three definitions have been used
interchangeably and have generally been considered equivalent and
consistent. But is this really the case?

Many diverse results show those views to be grossly oversimplified
(see, e.g., \citealt*{wgf97} for an early review). For example, there
is now much evidence that some bulges are really concentrated (thin)
discs, or equivalently that the central parts of discs can have a
density profile steeper than that of the outer disc (see, e.g.,
\citealt{k93} and \citealt{kk04} for reviews). The central surface
brightness profiles of most bulges also rarely approach the classic
$R^{1/4}$ law \citep[e.g.][]{apb95,bgdp03}. Although our own surface
brightness profiles are not ideal to tackle this last issue, because
of the line-of-sight integration through the outer disc, many objects
are nearly exponential in the center and we specifically study the
minor-axis profiles in \citeauthor{aab06}. Most importantly, our
sample is ideal to test the equivalence of thick structures with steep
light profiles and high velocity dispersions.

The images and surface brightness profiles of Figure~\ref{fig:bps}
reveal that, in some galaxies, the thick central structure (defined
isophotally) extends past the steep inner region of the surface
brightness profile (both along the major-axis and vertically
summed). Those two definitions of a bulge are thus neither equivalent
nor consistent, at least in those cases. Our observations thus show
that the steep inner region of the surface brightness profile is often
contained {\em within} the thick central component. This was already
hinted at by the poorer optical profiles of \citet{cb04}, although
\citet{ldp00b} found them to be generally equal. Examples include
ESO~151-G004, NGC~2788A and ESO~443-G042, but there are many. In fact,
our observations suggest that those two bulge definitions only seem to
agree systematically in galaxies with a boxy-shaped (or round) bulge,
such as NGC~1381 and IC~5096, while the disagreement is worst in some
of the the strongest peanut-shape bulges. There is no straightforward
explanation for this in classic bulge formation scenarios, but it is a
natural consequence of bar-buckling/thickening mechanisms, where it is
related to the bar viewing angle (seen respectively end-on and
side-on).

Having argued that the steep inner region of the surface brightness
profiles is generally contained within the thick part of the galaxies,
it is also true that the intermediate and often flat region of the
surface brightness profiles always extends {\em beyond} the thick
component. Some of the best examples are found in the galaxies listed
above, but again there are many (see, e.g., NGC~1886 and
NGC~4469). This behaviour is hard to reconcile with classic bulge
formation scenarios, but it arises naturally in
bar-buckling/thickening ones. Indeed, while the (projected) surface
brightness profile at intermediate radii remains shallow or flat
within most of the barred region of a galaxy (i.e.\ over the entire
deprojected bar length, no matter what the viewing angle is), only
part of the bar is thick. So, while the fact that {\em most} thick
components are shorter than the flat intermediate region of the
surface brightness profile is simply a consequence that most bars are
not seen exactly side-on, the fact that {\em all} thick components are
shorter than the flat intermediate region reflects a fact much less
widely appreciated, i.e.\ that only a fraction of the bar is actually
thick. This point has not been stressed in the literature, but it is
nicely illustrated and discussed in some depth in \citet{a05}. It is
explained by the orbital structure of barred discs
\citep[e.g.][]{spa02a,spa02b,psa02} and is in fact visible in most
published $N$-body models \citep[e.g.][]{cdfp90,am02,a05}.
\subsection{Three-dimensional structure of B/PS bulges}
\label{sec:disc_3d}
\begin{figure*}
\begin{center}
\includegraphics[width=0.9\textwidth]{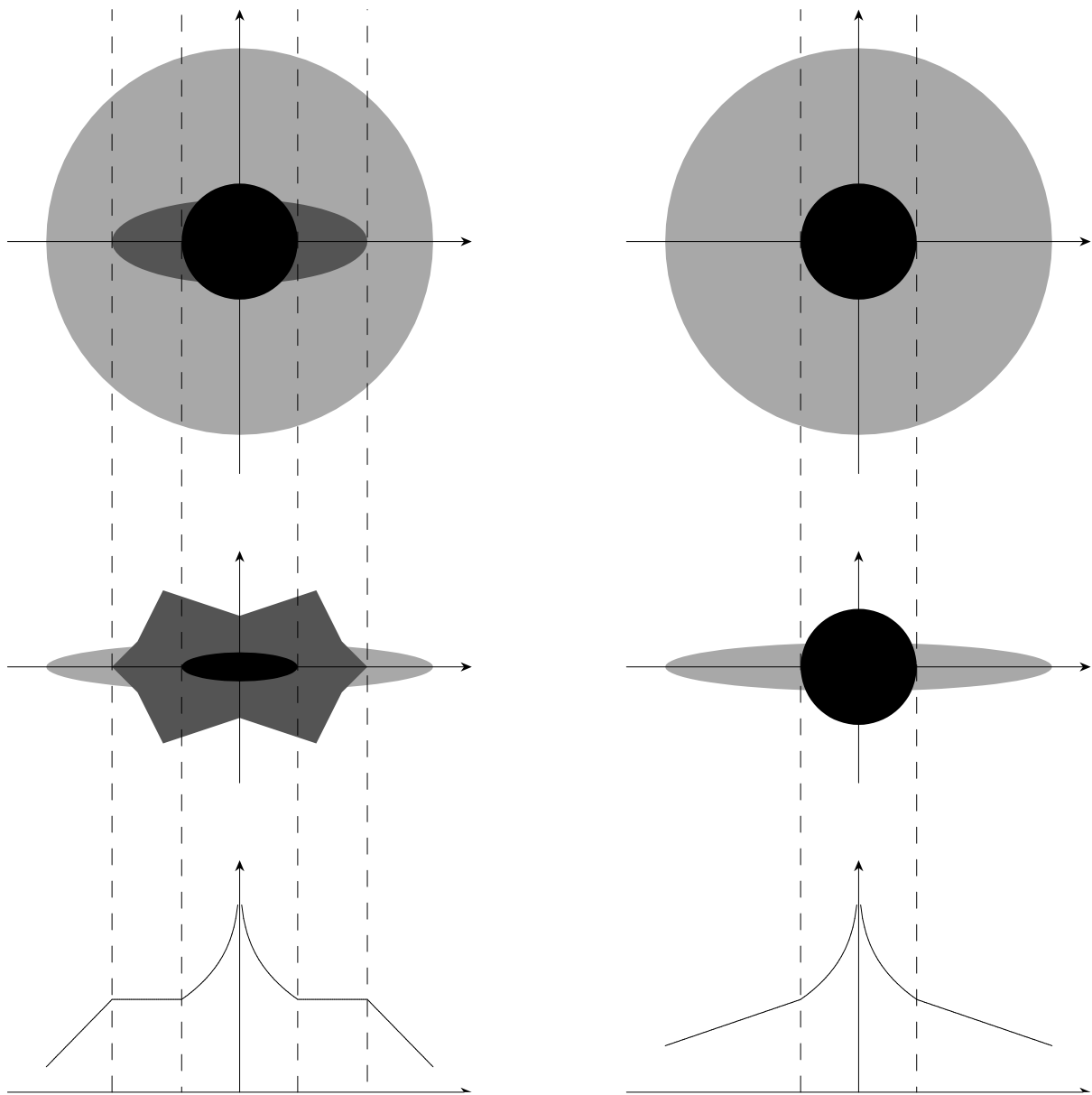}
\end{center}
\caption{Cartoon description of our B/PS bulge model (left) and the
  classic bulge model (right). From top to bottom, each panel shows
  the face-on view of the model, the edge-on (side-on) view and the
  edge-on surface brightness profile, all spatially
  registered. Galaxies with a classic bulge are generally viewed as
  having both a thick and steep inner (classic bulge) component within
  a thin exponential disc. We suggest that galaxies with a B/PS bulge
  instead harbour a thin and steep inner disc component (disc-like
  bulge) within a thick and shallow bar (B/PS bulge), followed by an
  outer exponential disc.}
\label{fig:bulges_cartoon}
\end{figure*}

A comparison of the major-axis and vertically-summed surface
brightness profiles reveals another interesting fact, i.e.\ that the
steep inner region of the surface brightness profiles is
systematically more pronounced and dominant in the major-axis profiles
than in the summed ones \citep[see also][]{ldp00b}. This is a key
indication of the 3D structure of galaxies with a B/PS bulge. It shows
that most of the material found at high $z$ belongs to the flat
intermediate region of the surface brightness profiles rather than to
the inner steep component, and that the latter has a smaller
characteristic height than the former (this will be shown more
transparently in \citeauthor{aab06}). Thus, the steep inner region of
the surface brightness profiles truly appears to be a thin
concentrated disc, while the shallow intermediate region appears to be
thick (and, according to the aforementioned arguments,
barred). Figure~\ref{fig:bulges_cartoon} presents a cartoon version of
this model, contrasted to the classic one.
%
%

\citet{a05} recently proposed a new nomenclature to differentiate
different types of bulges based on their formation mechanism. In
addition to classic bulges, which presumably form rapidly through
either collapse or merging, she identifies B/PS bulges, which are
simply (part of) thick bars, and disc-like bulges, formed by
(bar-driven) gaseous inflow and subsequent star formation. As more
than one type of bulge can coexist in a single object, the latter two
types are well suited to a description of the observations presented
here.

As noted above, \citet{k93} and \citet{kk04} review much of the
evidence for the interpretation of some bulges as concentrated discs,
which they include in a broader category of `pseudo'-bulges, generally
formed secularly (i.e.\ slowly). This vision of disc-like bulges does
not explicitely include the extended bulges discussed in this paper,
nor the many bulges which are isophotally thick and kinematically
hot. The thick bar model advocated in \citet{a05} and here addressess
all three points simultaneously. In this respect, we know that the
radial component of the velocity dispersion of the Milky Way bulge is
similar to that of its inner disc ($\approx100$~km~s$^{-1}$; e.g.\
\citealt*{sjw92,lf89}), and that the velocity dispersion of the
galactic bar/bulge is nearly isotropic. So although little is known
about either, a difference between the vertical velocity dispersion of
the inner disc and that of the bar/bulge must underlie their large
structural differences. While it is unlikely that concentrated discs
are hot vertically, (thick) bars appear to be
\citep[e.g.][]{gs05,dcmm05}.

One could argue that our model comprises a number of distinct building
blocks, like the classic models, but those blocks are very different
and tightly intertwined dynamically. First, the large-scale bar leads
to the formation of a concentrated disc, as in \citeauthor{k93}'s
(\citeyear{k93}) picture, but this disc is thin, largely decoupled
from the large-scale bar, and only addresses the first of the bulge
definitions (steep surface brightness profile). Second, the
large-scale bar itself is thick over most but not all its length, and
it addresses the second definition of bulges (isophotally thick
structure). As for the third definition of bulges (kinematically hot),
it is unclear which of the inner disc or large-scale bar dominates the
large velocity dispersions observed by \citet{cb04} in the equatorial
plane, but the models suggest that the culprit is slightly elongated
\citep{ba05}. As mentioned above, face-on bars do appear to have large
vertical velocity dispersions \cite[e.g.][]{gs05}.

The 3D structure proposed for B/PS bulges in \citet{a05} and here is
thus diametrically opposed to the classic one. Instead of a thick and
steep inner (classic bulge) component within a thin exponential disc,
we propose a thin and steep inner disc component (a disc-like bulge)
within a thick and shallow bar (a B/PS bulge), followed by an outer
exponential disc (see Fig.~\ref{fig:bulges_cartoon}). Furthermore,
while the two components of classic bulges are generally thought to
have formed separately through vastly different mechanisms and to
interact little with each other, the three `components' of galaxies
with B/PS bulges are largely made of the same material, which has
evolved and been shaped by the single dominant influence of the bar,
through the weak but relentless action of both radial and vertical
resonances. We do not appear to need additional classic bulges to
explain our observations, although it is difficult to rule them out
completely at the very center of our galaxies (i.e.\ on a scale much
smaller than the isophotally thick part).

We note that while this may seem to imply that no stellar population
gradient should exist in barred galaxies, this is not true. The radial
resonances, once established, will clearly affect the behaviour of the
gaseous (dissipative) material, which in turns feeds star
formation. It is thus expected that a certain segregration of the
stellar populations (mainly due to influence of the radial resonances)
will develop over time. While pure $N$-body models do not capture this
richness, others models including gas dynamics and star formation do
(e.g.\ \citealt*{fbk94,hs94,fb95}; but see also
\citealt{nn03}). However, the current simplicity of those models and
the ignorance of much of the physics involved (primarily regarding
star formation) prevent a detailed comparison with current
obsvervations, which are in any case limited \citep*[but
see][]{mr94,zkh94,ffi96}. There is no doubt, however, that the next
step for bar-driven bulge formation models is to include realistic
star formation and stellar population information (chemodynamics; see,
e.g., \citealt*{sht97,sbhtafj04,mw04}).

The fact that the maximal radial extent of the inner disc appears to
be systematically equal to or slightly smaller than the radius where
the bar is thickest (see Fig.~\ref{fig:bps}) may reflect some form of
`self-regulating' dynamics, i.e.\ that the inner disc is limited by a
radial resonance (perhaps the radial inner Lindblad resonance, ILR)
coinciding with a vertical resonance (perhaps the vertical ILR), as
advocated e.g.\ by \citet{cdfp90}. This should be explored further.
\subsection{Control sample}
\label{sec:disc_control}
Due to its limited size compared to the number of galaxies with a B/PS
bulge in our sample, the control sample was always meant more for
qualitative checks than for statistically robust comparisons. However,
it is now clear that some galaxies do not belong in it. For example,
\citet{ldp00a} classified the bulge of NGC~3957 as close to
boxy-shaped, \citet{bf99} and \citet{cb04} both found kinematic
indications of weak bars in NGC~3957 and NGC~4703, and \citet{cb04}
also found indication of non-axisymmetric motions in NGC~7123. The
current $K$-band imaging also shows centered X features in NGC~3957
and NGC~7123 and secondary maxima in NGC~3957 and NGC~4703. All three
galaxies show at least $3$ regions in their surface brightness
profile. The contrast between the B/PS bulge and control samples would
thus be greater if NGC~3957 and NGC~4703 (and perhaps NGC~7123) were
appropriately classified as B/PS.

Even if bars are peanut-shaped, a fraction of round bulges should
still be expected to be bars, as the latter appear round when seen
end-on. This fraction is however expected to be small, since already
for an angle of $10$--$30\degr$ between the bar major-axis and the
line-of-sight a boxy shape is observed
\citep[e.g.][]{cdfp90,ldp00b,a05}. The large contamination of our
control sample thus probably arises from the way it was
selected. Indeed, no catalog of edge-on galaxies with a round bulge
existed when it was constructed, so the control sample galaxies were
selected from apparently misclassified objects in catalogs of galaxies
with a B/PS bulge \citep{j86,sa87,s87} and from the Flat Galaxy
Catalog of \citet{kkp93}. Carrying a study along the lines of the
current one and those of \citet{bf99} and \citet{cb04}, but for an
enlarged and properly selected sample of round bulges \citep[for
example using the more recent catalog of][]{ldp00a}, would thus be
valuable. While this represents a significant amount of work, showing
with more certainty that round bulges behave differently from B/PS
bulges would greatly strengthen our conclusions.
%
%
\section{SUMMARY AND CONCLUSIONS}
\label{sec:conclusions}
We have presented \kn-band imaging observations of a sample of $30$
edge-on spiral galaxies, most of which harbour a boxy or peanut-shaped
(B/PS) bulge. Those data are minimally affected by dust and best trace
population II stars, where most of the luminous mass resides. Our
multi-faceted analysis suggests that B/PS bulges are simply the thick
part of bars viewed edge-on (see Figures~\ref{fig:bps} and
\ref{fig:control}).

Galaxies with a B/PS bulge tend to have a more complex morphology than
galaxies with other bulge types, more often showing centered or
off-centered X structures, secondary maxima along the major-axis, and
spiral-like structures. Best revealed by unsharp-masking, those
features are also observed in three-dimensional N-body simulations of
barred discs \citep{a05}, and can be explained by the orbital
structure of bars \citep[see, e.g.,][]{psa02}, although they need not
be uniquely related to them. Only minor-axis extrema may be
preferentially related to other bulge types. Whether taken along the
major-axis or summed vertically (to simulate a flat galaxy), the
surface brightness profiles of galaxies with a B/PS bulge are also
more complex, more often showing $3$ or more clearly separated
regions, including a rather shallow or flat intermediate region (see
Figure~\ref{fig:sbprof_cartoon}). Such Freeman Type~II profiles are
expected from barred galaxies \citep[e.g.][]{ba05}, but they do not
have a natural self-consistent explanation in classic bulge formation
scenarios.

The radial breaks observed in the vertically-summed profiles of our
objects provide further evidence of the transfer of angular momentum
and radial redistribution of disc material mediated by the (presumed)
bars \citep[see, e.g.,][]{a02,a03}. Furthermore, the spatial
correlations of the radial breaks with the ionized-gas and stellar
kinematics \citep{bf99,cb04} are as expected for fast bars, currently
favoured by observations. The differences between the major-axis and
vertically-summed profiles provide evidence for abrupt variations of
the scaleheight of the disc material. This is, again, as expected from
the diverse orbital families and vertical resonances and instabilities
present in barred discs, but contrary to conventional wisdom. A
quantitative and robust analysis of those scaleheight variations and a
comparison with $N$-body simulations will appear in future papers of
this series.

Three other facts stand out. First, the steep inner region of the
surface brightness profiles is systematically equal to or shorter than
the isophotally thick part of the galaxies. Second, the isophotally
thick part is itself systematically contained within the flat
intermediate region of the surface brightness profiles. Third, the
steep inner region of the surface brightness profiles is much more
prominent along the major-axis than in the vertically-summed profiles.

We are thus led to radically alter the classic `bulge + disc' model,
composed of a thick and steep spheroidal bulge largely decoupled from
a thin (possibly barred) exponential disc. Analogously to \citet{a05},
we propose here that galaxies with a B/PS bulge are composed of a thin
concentrated disc (a disc-like bulge), formed secularly by the bar and
responsible for the steep inner region of the surface brightness
profiles, contained within a (partially) thick bar (the B/PS bulge),
responsible for the flat intermediate region of the surface brightness
profiles and the complex morphological structures, itself contained
within a thin outer exponential disc (see
Figure~\ref{fig:bulges_cartoon}). Those components are closely
intertwined dynamically and are largely made of the same (disc)
material, shaped over long timescales by the bar.

The challenge to any competing formation scenario for galaxies with a
B/PS bulge, which represent at least $45$~per cent of the local galaxy
population \citep{ldp00a}, is thus to simultaneously and
self-consistently explain, equally well or better, their numerous
morphological, photometric, and kinematic properties, as well as the
correlations between them.
%
%
\section*{Acknowledgments}
The authors would like to thank P.\ Patsis and J.\ Kormendy for useful
discussions, and M.\ Pohlen and D.\ Vergani for technical help. MB
acknowledges support from NASA through Hubble Fellowship grant
HST-HF-01136.01 awarded by Space Telescope Science Institute, which is
operated by the Association of Universities for Research in Astronomy,
Inc., for NASA, under contract NAS~5-26555, during much of this
work. MB also acknowledges the hospitality of the Universit\'{e} de
Montr\'{e}al and Nagoya University while preparing the
manuscript. This project took shape partly through discussions during
a Guillermo Haro workshop. The participation of EA and AB to this
workshop was facilitated by an ECOS/ANUIEF exchange grant. We
thank the staff of Mount Stromlo and Siding Spring Observatories for
their assistance during and after the observations. This project made
use of the LEDA database: {\tt http://leda.univ-lyon1.fr/}. The
NASA/IPAC Extragalactic Database (NED) is operated by the Jet
Propulsion Laboratory, California Institute of Technology, under
contract with NASA. The Digitized Sky Surveys were produced at the
Space Telescope Science Institute under U.S.\ Government grant
NAG~W-2166. The images of these surveys are based on photographic data
obtained using the Oschin Schmidt Telescope on Palomar Mountain and
the UK Schmidt Telescope. The plates were processed into the present
compressed digital form with the permission of these institutions.
%
%

%
\end{document}